\documentclass{aa}  

\usepackage{graphicx}
\usepackage{txfonts}
\usepackage{lscape}
\usepackage{tabularx}
\usepackage{natbib}
\usepackage{caption}
\begin{document}

\newcommand{\oiii}{[O{\,\scshape iii]}}
\newcommand{\sii}{[S{\,\scshape ii]}}
\newcommand{\nii}{[N{\,\scshape ii]}}
\newcommand{\ha}{H$\alpha$}
\newcommand{\hb}{H$\beta$}

   \title{Detection and characterization of detached tidal dwarf galaxies}


   \author{Javier Zaragoza-Cardiel
          \inst{1}
          \and
          Beverly J. Smith\inst{2}
          \and Mark G. Jones\inst{2}
          \and Mark L. Giroux\inst{2}
          \and Shawn Toner\inst{2}
          \and Jairo A. Alzate\inst{1}
          \and David Fern\'andez-Arenas\inst{3}
          \and Divakara Mayya\inst{4}
          \and Gisela Ortiz-León\inst{4}
          \and Mauricio Portilla\inst{4}
}

   \institute{Centro de Estudios de Física del Cosmos de Aragón (CEFCA)
, Plaza San Juan 1,\\
E-44001, Teruel, Spain\\
              \email{jzaragoza@cefca.es}
             \and 
             East Tennessee State University\\
Department of Physics and Astronomy\\
Johnson City, TN 37614, USA\\
              \email{smithbj@etsu.edu}
              \and
              Canada–France–Hawaii Telescope, Kamuela, HI 96743, USA
              \and
              Instituto Nacional de Astrofísica, Óptica y Electrónica, Luis Enrique Erro 1, Tonantzintla 72840, Puebla, Mexico
                 }

%


 
  \abstract
   {
Tidal interactions between galaxies often give rise to tidal tails, which can harbor concentrations of stars and interstellar gas resembling dwarf galaxies. Some of these tidal dwarf galaxies (TDGs) have the potential to detach from their parent galaxies and become independent entities, but their long-term survival is uncertain. In this study, we conducted a search for detached TDGs 
associated with a sample of 39 interacting galaxy pairs in the local Universe using infrared, ultraviolet, and optical images. We employed IR colors and UV/optical/IR spectral energy distributions to identify potential interlopers, such as foreground stars or background quasars. Through spectroscopic observations using the Boller and Chivens spectrograph at San Pedro M\'artir Observatory, we confirmed  that six candidate TDGs are at the same redshift as their putative parent galaxy pairs.  
 We identified and measured emission lines in the optical spectra and calculated nebular oxygen abundances, which range from log(O/H) = 8.10 $\pm$ 0.01 to 8.51 $\pm$ 0.02. We have serendipitously discovered an additional detached TDG candidate in Arp72 using available spectra from SDSS.
Utilizing the photometric data and the CIGALE code for stellar population and dust emission fitting, we derived the  stellar masses, stellar population ages, and stellar
metallicities for these detached TDGs. 
 Compared to standard mass-metallicity relations for dwarf galaxies, five of the seven candidates have higher than expected metallicities, 
confirming their tidal origins. One of the seven candidates remains unclear due to large uncertainties in metallicity, and another  has stellar and nebular metallicities compatible with those of a preexisting dwarf galaxy. The latter object is relatively compact in the optical relative to its stellar mass, in contrast to the other candidate TDGs, which have large diameters for their stellar masses compared to most dwarf galaxies. 
The derived stellar population ages range from 100 Myr to 900 Myr, while the inferred stellar masses are between 2 $\times$ 10$^6$~M$_{\sun}$ and 8 $\times$ 10$^7$~M$_{\sun}$.  
Four of the six TDGs are associated with the gas-rich M51-like pair Arp 72,  one TDG is associated with a second M51-like pair Arp 86, and another is associated with Arp 65, an approximately equal mass pair.
In spite of the relatively low stellar masses of these TDGs, they have survived for at least 100 - 900 Myrs,
suggesting that they are stable and in dynamical equilibrium.
 We conclude that encounters with a relatively low-mass companion (1/10th $-$ 1/4th of the mass of the primary) can also produce long-lasting TDGs.}

   \keywords{galaxies: interactions --- 
galaxies: dwarf ---  galaxies: evolution --- galaxies: abundances --- galaxies: peculiar
               }

\titlerunning{Detached tidal dwarf galaxies}
\authorrunning{Zaragoza-Cardiel et al.}
               
   \maketitle
%

\section{Introduction} \label{sec:intro}

Galaxy interactions can spawn tidal dwarf galaxies (TDGs), concentrations
of gas and young stars in tidal features that resemble 
dwarf galaxies
\citep{duc1994, duc1997, duc2012}.
Unlike most dwarf galaxies, TDGs are expected to lack a 
massive dark matter component
since they are made of material pulled out from the disk 
\citep{barnes1992,
duc2004,
bournaud2006,
2007MNRAS.375..805W,
lelli2015}.  
Thus, TDGs present a unique opportunity to explore the role of dark matter in galaxy dynamics and 
understand the interplay between baryonic matter and the invisible dark matter component.
 
However, simulations suggest dark matter 
may be tidally stripped from dwarf galaxies
during interactions with massive companions
\citep{jackson2021}. Therefore, a lack of
dark matter may not be a definitive signature of a TDG.

One way to potentially distinguish TDGs from preexisting dwarf galaxies is via their metallicities.
TDGs are expected to have higher abundances of heavier elements than other dwarf galaxies of the same mass
and optical luminosity if they were formed from gas and stars previously located within the disk of a more massive galaxy.
Hence, comparison to mass-metallicity and luminosity-metallicity relations for normal
dwarf galaxies has traditionally been used to identify TDGs \citep{duc2000, weilbacher2003a, croxall2009, duc2012, sweet2014}.
However, TDGs formed early in the history of the Universe may have low metallicities similar to other dwarfs \citep{hunter2000, recchi2015},
though this has been questioned \citep{2021MNRAS.503.2866D}.
Also, TDGs formed from metal-poor gas from the outskirts of galaxies have a low metallicity \citep{hunter2000}.

Sometimes TDGs can be misidentified as dark galaxies \citep{2021A&A...649L..14R}, which are 
predicted to be composed of mostly dark matter and devoid of significant stellar populations due 
to the lack of star formation in low-mass halos and at low metallicity \citep{2010ApJ...714..287G}. 
Distinguishing between TDGs and dark galaxies requires careful investigation of 
the dynamics, stellar populations, environment, and history of a system 
\citep{2015AJ....149...72C,2021A&A...649L..14R,2024A&A...681A..15M,2024ApJ...964...85D}.

Some TDGs can detach from their parent galaxies and become independent
dwarf galaxies 
\citep{zwicky1956},
but how frequently this actually
happens is a matter of debate. 
Some estimates suggest that the majority of currently observed
dwarf galaxies may have been tidally formed rather than being primordial
\citep{hunsberger1996, okazaki2000, dabringhausen2013}.
Other calculations indicate that $\le$10\% of local dwarfs originated
as TDGs
\citep{bournaud2006, bournaud2010, wen2012, kaviraj2012}.
The reason for the difference of opinion is that the TDG formation rate and
survival timescale are not well determined.
It has been suggested that many of the dwarf galaxies in the Local
Group were formed tidally 
\citep{lynden1976, metz2007, pawlowski2012, fouquet2012, hammer2013,
yang2014, pawlowski2018};
however, this idea has been
questioned 
\citep{duc2014, collins2015, cautun2015,2021MNRAS.504.1379S,2023NatAs...7..481S,2023MNRAS.520.3937P}.

Factors that affect the longevity of TDGs include 
mass loss due to stellar winds and supernovae 
\citep{ferrara2000, recchi2007, recchi2013, ploeckinger2015};
tidal disruption and compression 
\citep{bournaud2006, ploeckinger2015};
and infall back into the parent galaxy 
\citep{bournaud2006, hancock2009}.
If the star formation rate (SFR) in the TDG is high, stellar feedback may 
eject much of the gas from the TDG, hastening its demise.
Ram pressure stripping by intracluster or intragroup gas or hot
halo gas may also affect the evolution of a TDG 
\citep{smith2013}.
Detailed hydrodynamical simulations 
\citep{bournaud2006}
indicate that more massive objects are more likely to survive,
particularly TDGs on low eccentricity orbits formed near the ends of
tidal tails and objects produced by approximately equal-mass galaxy pairs.
The size and density of satellite dwarfs also affects survival rates
\citep{casa2012}.
Interactions and mergers of gas-rich galaxies, such as those at
high redshift, produce more TDGs on average 
\citep{bournaud2011, fouquet2012}.

Even if TDGs do not survive as independent galaxies, their temporary
existence may affect the evolution of their parent galaxy and its 
environment. Star formation and feedback in TDGs may enrich 
halo gas as well as the intergalactic medium 
\citep{duc2014}.
Dispersion of the 
star clusters and individual stars in TDGs and tidal features 
may contribute to the intragroup, intercluster, and galactic halo 
globular clusters and diffuse starlight \citep{mihos2005}.
Further, the infall of tidal debris back onto a parent disk 
\citep{hibbard1995}
may 
help build the bulge or trigger star formation in the disk.

To better understand the contributions of TDGs to galaxy
evolution, it is necessary to look for those that are the most evolved. One way to approach this is to look for 
detached TDGs,
objects 
that are no longer connected to their parent galaxies by
tidal structures. 

To this end,
we have conducted a systematic multi-wavelength search for detached 
TDGs in the vicinity of a well-defined sample of 
nearby interacting galaxies. 
In this study, we have searched for possible
star-forming TDGs 
outside of optically bright tidal features. We confirmed 
the redshift coincidence of
six candidates via spectroscopic observations. 
We derived ages, metallicities, and stellar masses for these sources in order 
to constrain the timescale  of  formation and  their evolutionary history, which we have compared with theoretical models of TDG formation and evolution.

In Section 2, we describe our sample of interacting galaxy pairs and the available 
imaging data.
In Section 3, we explain how we selected candidate TDGs; determined 
their fluxes in UV, optical, and IR bands; and removed possible interlopers.
We present the optical spectroscopy observations  and show how we obtained redshifts in Section 4. In Section 5, 
we provide detailed descriptions of the systems with spectroscopically confirmed 
TDGs, and we present 
the analysis of the combined datasets in Section 6. 
In Section 7, we present the results and compare them to dwarf galaxy
relations. We present a discussion of the results and compare with other TDGs in Section 8. Conclusions are given in 
Section 9.

\section{Galaxy samples and data} \label{sec:sample}

\subsection{Samples}

We started our search for TDGs with the
46 nearby ($<$150 Mpc) strongly interacting 
pre-merger pairs of galaxies chosen from the Arp Atlas, the "Spirals, Bridges,
and Tails" (SB\&T) sample \citep{smith2007, smith2010, smith2016, smith_struck2010, zaragoza2018}.
This sample was previously used to compare 
the global SFRs and specific SFRs (SFR/stellar mass)
of interacting versus spiral galaxies 
using IR and UV photometry
\citep{smith2007,smith2010, smith_struck2010,
zaragoza2018}.  
The properties of individual star-forming regions within the disks
and tidal features
were studied by 
\cite{smith2016} and \cite{zaragoza2018}.
In the current study, we search for detached TDG candidates outside
of optically bright tidal features.

To improve the reliability of our search for TDGs,
to be included in our final
list of TDG candidates,
we required objects to be 
detected in both the near-ultraviolet (NUV) and at 8 $\mu$m (see Section 3.1).
Furthermore, since we used \textit{Spitzer} infrared colors to 
eliminate foreground stars and background quasars from our sample
(see Section 3.3),
we required the field to be imaged by \textit{Spitzer} in the
3.6 $\mu$m and 4.5 $\mu$m filters as well.
This decreased our  initial sample to the 40 interacting pairs that 
have 
3.6 $\mu$m, 4.5 $\mu$m, and 8 $\mu$m and NUV images available.
Closer inspection of the available images showed that one of the 
remaining SB\&T galaxies has a field of view of little use
around the galaxies. This system was 
therefore 
removed from the sample,
leaving a final sample of 39 SB\&T pairs.
 These galaxies have a median distance of
44 Mpc.

 These interacting galaxies 
are more nearby than many used in previous searches
for TDGs.
For comparison,
the SDSS-selected TDGs studied by 
\cite{kaviraj2012}
have a median distance of 226 Mpc, the 
\cite{hunsberger1996}
TDGs
in compact groups have a median distance of 92 Mpc,
and 
the  (ultra) luminous infrared galaxies (U/LIRGs) searched by 
\cite{miralles2012}
for TDGs have a median distance of 159 Mpc.
Our galaxies are also closer than the compact groups surveyed
by 
\cite{eigenthaler2015}
for TDGs, which have a typical 
distance of 123 Mpc.
This means that our survey 
will be more sensitive 
to lower mass objects.
However, our galaxies are more distant on average than the TDGs
studied in the 
spectroscopy 
survey of 
\cite{duc2014},
which are all closer
than 40 Mpc.  
Our galaxies are also more distant on average
than the 
merger remnants surveyed for dwarf companions 
by 
\cite{delgado2003},
which have a median distance
of about 36 Mpc. 

Most previous surveys of TDGs focused on objects 
still embedded in the optically bright tidal structures of interacting
galaxies (e.g., 
\citealp{duc1994, duc1997, smith2010}),
while
in the current study we searched for star-forming objects outside
of classical tidal features.    Another difference from earlier surveys
is that our systems are relatively isolated pre-merger pairs, 
in contrast to TDGs in compact
groups 
\citep{hunsberger1996, eigenthaler2015},
TDGs associated with
post-merger early-type galaxies 
\citep{duc2014},
or merger remnants and late-stage mergers 
\citep{delgado2003}.
In the current study, we focus on the evolutionary
stage between tidal tail formation and final galaxian merger.
 Understanding the dynamical history of
these systems will be easier than for later-stage mergers or
compact groups.
Since our sample is optically selected, the galaxies
are less biased toward extreme starbursts than the LIRG/ULIRG
systems
studied by 
\cite{miralles2012}.

\subsection{ Imaging data} 

Most of the SB\&T systems have been observed in the IR 
at 3.6, 4.5, 5.8, 8.0, and 24 $\mu$m by the \textit{Spitzer} telescope, 
and in the far-UV (FUV) and near-UV (NUV) filters
by the Galaxy Evolution Explorer (GALEX) satellite
(Smith et al.\ 2007; 2010; Smith et al.\ 2016).
In the current study, we only used GALEX images with exposure
times greater than 1000 seconds.
Many of the SB\&T galaxies also have Sloan Digitized Sky Survey (SDSS)
optical {\it ugriz} images available 
\citep{smith2010, smith2016}.
 For systems without SDSS images, 
we use Dark Energy Spectroscopic Instrument  Legacy Imaging Surveys (DESI-LIS) {\it grz} images when available 
\citep{dey2019}.
About half of the SB\&T galaxies have narrowband 
H$\alpha$ maps available 
\citep{smith2016}.
About one third have been mapped in the 21 cm HI line.
We also used Two Micron All-Sky Survey (2MASS) 
J, H, and K images\footnote{https://www.ipac.caltech.edu/2mass/overview/about2mass.html} to obtain near-IR
photometry of our candidate TDGs.

\section{Candidate tidal dwarf galaxies} \label{sec:TDGs}

\subsection{Selection of candidate tidal dwarf galaxies} 

In an earlier study of 
the SB\&T interacting sample 
\citep{smith2016},
we searched for TDGs and knots of 
star formation associated with 
tidal structures brighter than 
an SDSS g band surface brightness $\mu$$_{\rm g}$ of 24.58 mag~arcsec$^{-2}$ 
(equivalent to a B band surface brightness $\mu$$_{\rm B}$ = 25 mag~arcsec$^{-2}$).
When SDSS images were not available, we use an NUV surface
brightness of 26.99 mag~arcsec$^{-2}$, which is the equivalent
surface brightness assuming a typical NUV $-$ g color for
tidal tails 
\citep{smith2016}.

We now extend that procedure to star-forming regions 
outside of those limiting isophotes.   In this study, we define a "detached TDG" as a TDG that is not connected to
the parent galaxy by an optical bridge brighter than $\mu$$_{\rm g}$
= 24.58 mag~arcsec$^{-2}$.   These may be TDGs that have already
detached from their parent tidal tail or bridge, or whose host tidal
structure has stretched or dispersed sufficiently to produce lower surface
brightnesses.
 Our definition of a detached TDG is strictly an operational
definition, based on observed optical surface brightness, and is
independent of whether or not the TDG is gravitationally bound to
the parent galaxy.
These candidate detached TDGs may have
formed during the current interaction, or they may have been produced
during an earlier close encounter between the galaxies in the pair.
In some cases, two galaxies in a pair may undergo multiple close encounters
before a final merger 
(e.g., \citealp{hancock2007, dobbs2010}).
Thus more than one generation of TDGs may be produced.

We use the \textit{Spitzer} 8 $\mu$m images to identify knots of star 
formation that lie outside of the SDSS g band 24.58 mag~arcsec$^{-2}$ 
isophote or its equivalent.
Since the field of view of the GALEX images is large (1.2 degrees diameter),
the primary factor that limits how far from the galaxy we can search 
is the \textit{Spitzer} 8 $\mu$m field of view, which is $\ge$5$'$ for these galaxies,
corresponding to galactic radii of 55 kpc $-$ 200 kpc.
Computer simulations of TDG formation indicate that the majority of
TDGs are found within this distance of the progenitor 
\citep{bournaud2006}.

Our 8 $\mu$m plus NUV selection criteria  biases the sample
toward regions that have formed stars in the last 100 Myrs (e.g.,
\citealp{kennicutt2012}).
Thus, we may miss older sources
 without at least a subset of relatively young stars. However,
detection in both of these bands will improve reliability,
better ensuring that the
sources are star-forming regions rather than foreground
stars or background quasars (see Section 3.3).
These objects may be knots of star formation that originated in the galaxies
and then detached,  or dense clumps of gas in extended gas-rich tidal features
that have gravitationally collapsed and formed stars.

The TDG selection process was done automatically, searching 
the 8 $\mu$m maps for 
star-forming regions
on two different spatial scales, 1 kpc and 2.5 kpc.
Knots of star formation within galaxy disks and tidal features are typically of 
this size 
\citep{smith2005, smith2008, hancock2007, hancock2009},
and 
previously identified TDGs have radii of $\sim$1 $-$ 2 kpc
\citep{miralles2012, duc2014}.

To find candidate TDGs,
we smoothed each \textit{Spitzer} 8 $\mu$m
image
to a full width at half maximum (FWHM) resolution of 2.5 kpc,
based on the distance to the galaxy. 
We then used the IRAF\footnote{Image Reduction and Analysis Facility,  distributed by the National Optical Astronomy Observatories, which
are operated by the Association of Universities for Research in Astronomy,
Inc., under cooperative agreement with the National Science Foundation} 
routine {\it daofind} to search for sources 
on the smoothed image on this scale.   
With {\it daofind}, we used a detection threshold
of 10$\sigma$ above the noise level in the smoothed images.  
For the 30 galaxy pairs 
closer than 67 Mpc, we repeated this process using a FWHM
of 1.0 kpc to search for objects on a smaller spatial scale.
The 67 Mpc distance
limit is set by the spatial resolution of the GALEX and \textit{Spitzer} 24 $\mu$m,
which limits reliable aperture photometry on smaller scales for more
distant galaxies (see \citealp{smith2016}).

The edges of many of the \textit{Spitzer} 8 $\mu$m images have poorer 
sensitivity due to 
decreased coverage during mapping observations.  This means that
the {\it daofind} results will be less reliable in these regions.
To eliminate these portions of the sky,
we used the \textit{Spitzer} 8 $\mu$m coverage maps to 
mask the parts of the images that have \textit{Spitzer} coverages less than 65\%
of values in the inner
regions of the galaxies.
To further avoid
edge effects we conservatively expanded the mask to include the portions 
of the
images just inside this 65\% coverage level, within one resolution
element on the smoothed image (i.e., within 2.5 kpc or 1.0 kpc).
Since we search for possible foreground or background objects 
based on the \textit{Spitzer}
[3.6] $-$ [4.5] and [3.6] $-$ [8.0] colors (see Section 3.3),
we also mask portions of the sky with less than 65\% coverage 
in the 3.6 $\mu$m
and 4.5 $\mu$m bands. 
Also, since we use detection in the NUV as an additional criteria,
we mask areas that are not covered in the NUV image.
For a few of the galaxies in our sample,
very large field of view 
\textit{Spitzer} images are available. 
For these systems, we masked all portions of the
images at distances greater than 1 Mpc from both galactic nuclei.
All candidate sources found in the masked areas were eliminated from
the survey.
If there are other known galaxies in the \textit{Spitzer} 8 $\mu$m field
at a different redshift from the target galaxy, we masked those galaxies
in the image.

\subsection{Photometry of candidate tidal dwarf galaxies} \label{sec:photometry}

 For each candidate TDG, we extracted photometry in all of the available
filters using the IRAF {\it daophot} command.
The photometry was done on the original unsmoothed images using
aperture radii of either 1.0 kpc or 2.5 kpc. 
We applied aperture corrections to the photometry to account for
spillage outside of the aperture due to the point spread function.
The aperture corrections were calculated
as in 
\cite{smith2016}
and \cite{zaragoza2018}.

\subsection{Identifying possible interlopers in the tidal dwarf galaxy sample } \label{sec:interlopers}

Some of  the candidate TDGs found by the above procedure may be foreground stars, background quasars,
or background galaxies.
 Before doing follow-up spectroscopic observations to confirm redshifts,
we first
searched the NASA Extragalactic
Database\footnote{http://ned.ipac.caltech.edu} (NED) 
 for published redshifts of the objects, and eliminated known background
objects.

In the absence of published redshift measurements, some interlopers can be
tentatively identified using their \textit{Spitzer} IR colors. 
As noted in 
Smith et al.\ (2016), sources
with [3.6] $-$ [8.0] $<$ 0.6 and $-$0.3 $<$ [3.6] $-$ [4.5] $<$ 0.3 
may be foreground stars  because in the 3.6-8$\mu$m range stars spectral energy distributions (SEDs) drop so [3.6]-[4.5] and [3.6]-[8.0] colors are near zero \citep{2004ApJS..154..315W} .  
Sources with [3.6] $-$ [4.5] $>$ 0.3 and 
1.0 $<$ [3.6] $-$ [8.0] $<$ 3.6 could be background active galactic nucleus (AGN)  because they are flat between 3.6-4.5$\mu$m range \citep{2005AJ....129.1198H} in contrast with the drop in this range of star-forming regions \citep{smith2005,smith2008,smith2014}, so background AGNs have redder [3.6] $-$ [4.5] colors.
Star-forming regions with metallicities larger than that of the Small Magellanic Cloud (SMC) \citep[$12+\rm{log}\frac{O}{H}=8.03$,][]{1992ApJ...384..508R} are unlikely to fall in these ranges \citep{smith2016}. Since TDGs are more metal rich than common dwarf galaxies we probably do not lose TDGs using these color criteria.

It is possible that low metallicity dwarf galaxies may
fall in these regions in color-color plots,
since galaxies with low metallicities
are often faint in the polycyclic aromatic hydrocarbon (PAH) features
that contribute to the \textit{Spitzer} 8 $\mu$m band 
\citep{houck2004, madden2006, hunt2010}.
This means that low metallicity objects 
may have bluer [3.6] $-$ [8.0] colors than higher metallicity galaxies
\citep{rosenberg2008}.
Furthermore, low metallicity systems may have redder [3.6] $-$ [4.5]
colors than solar-metallicity galaxies 
\citep{smith2009}.

 However, as noted above, true TDGs are expected to have higher metallicities 
than classical dwarfs and are thus expected to have \textit{Spitzer} colors similar to those
of star-forming regions within galaxy disks.  
We therefore eliminated candidates outside of the IR color ranges given above as likely
background or foreground objects.

 We also used published redshifted narrowband H$\alpha$ images of the systems (references tabulated
in Smith et al.\ 2016) to search for 
possible H$\alpha$ detections of the candidate TDGs.  
A reliable detection of a candidate TDG means that it is likely at the same redshift
as the parent galaxy. However, artifacts in the H$\alpha$ maps due to
imperfect continuum subtraction because of mis-matches in the point spread
function of the on-line and off-line images may lead to false detections at low S/N.
Thus tentative detections of H$\alpha$ from these images must be confirmed with 
follow-up spectroscopy.

 We also searched the literature for 21 cm HI maps of the systems,
and prioritized candidate TDGs that lie within or near HI concentrations
at the same redshift as the galaxy pair. Positional coincidence of a UV/optical/IR 
source with such an HI cloud is strong evidence that it may be associated with
the galaxy pair. However, it is possible that the UV/optical/IR source
is merely a background object behind the HI cloud, and thus
the redshift must be confirmed via optical spectroscopy.

\section{Longslit spectra observations}

\begin{figure}
\centering
\includegraphics[width=0.9\linewidth]{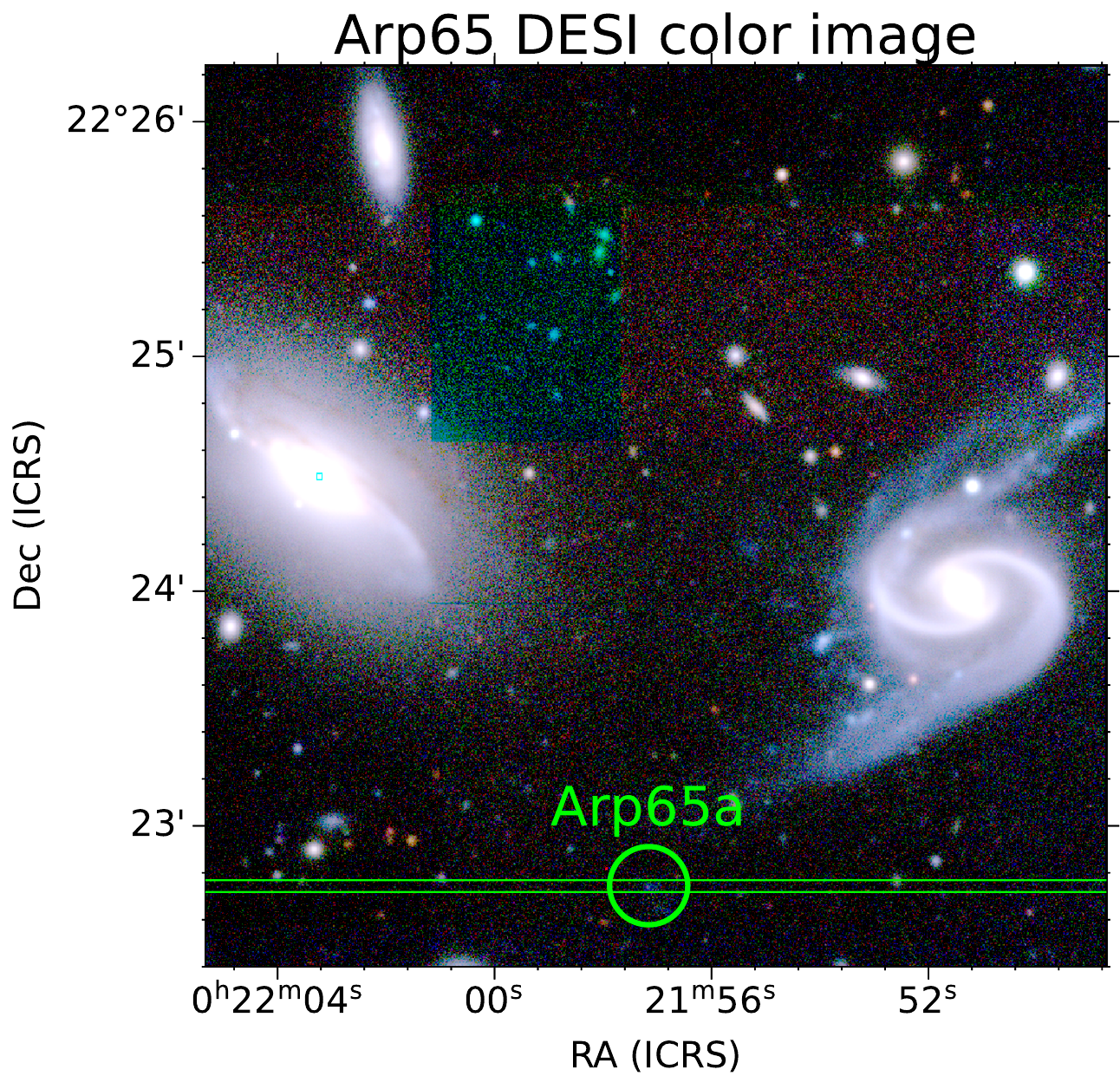} \\ 
\includegraphics[width=0.9\linewidth]{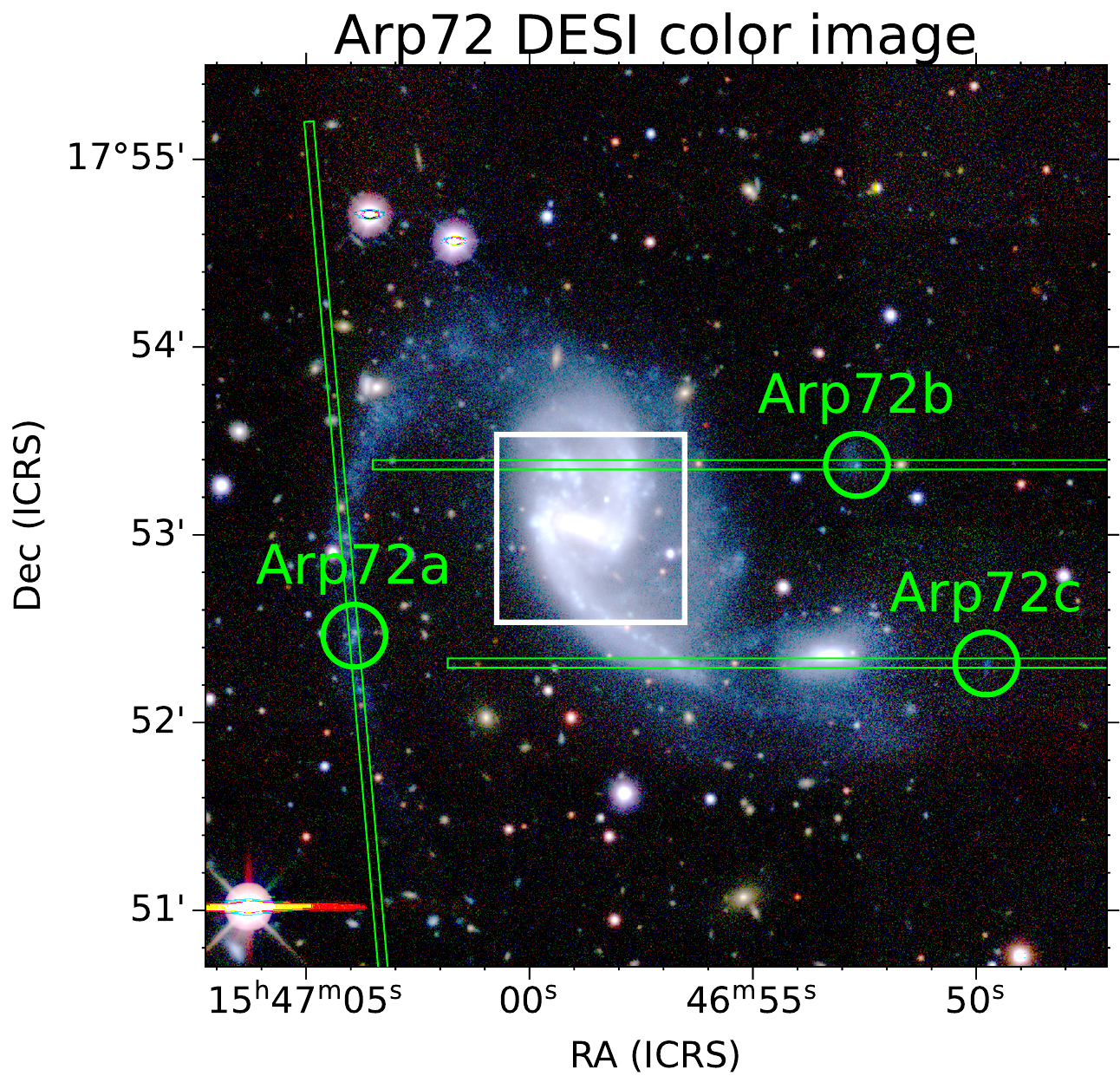} \\
\includegraphics[width=0.9\linewidth]{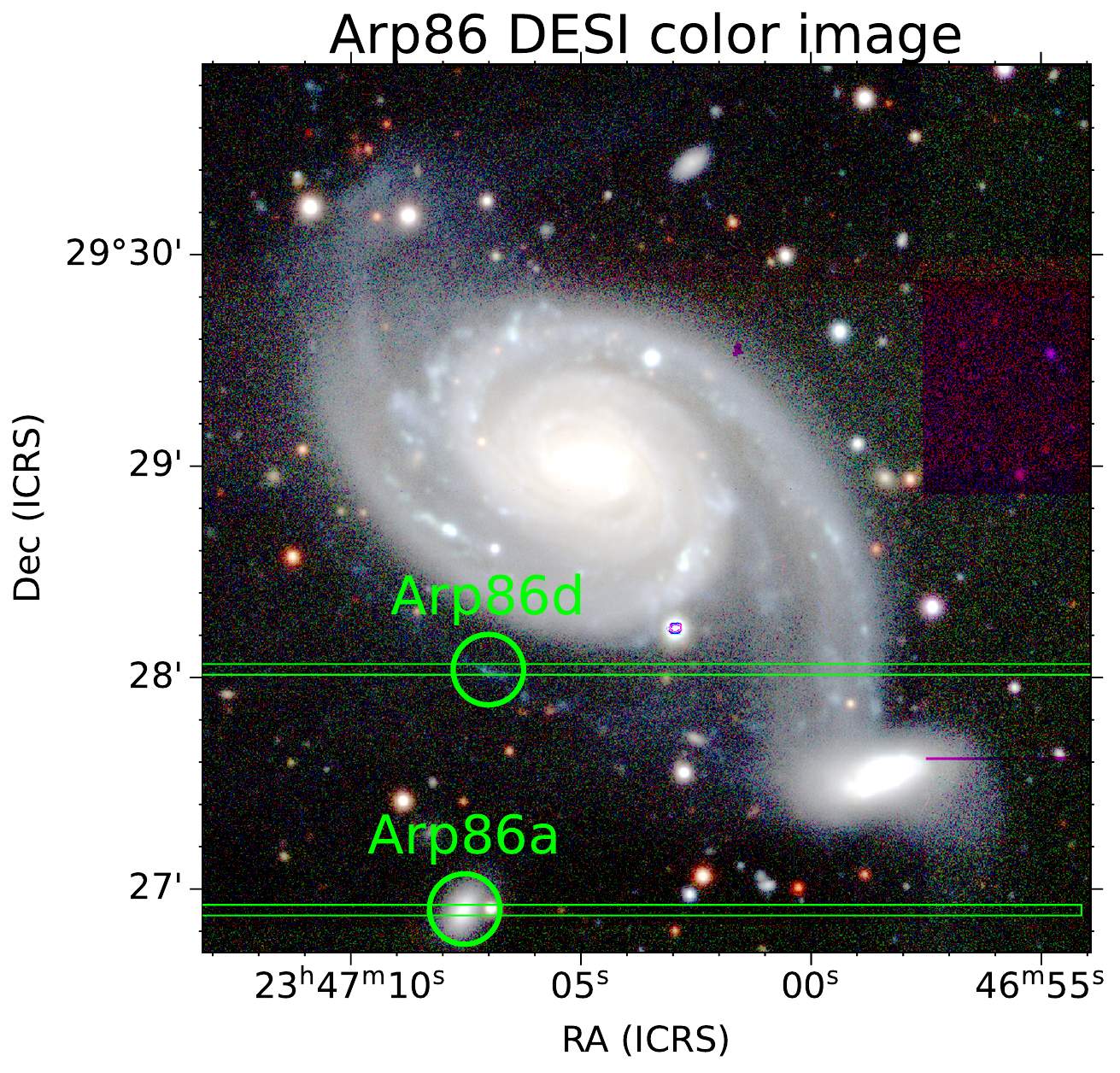} \\
\caption{\label{fig_obs}  DESI-LIS color image of the galaxies in which we have identified and confirmed detached TDGs. The TDGs are shown as green circles. The slit positions of the 
spectrograph are shown in green. The MUSE field of view for the primary
galaxy in Arp 72, NGC 5996, is marked by a white box in the middle
panel. The green square in the top panel is an artifact in the DESI-LIS color map.
}
\end{figure}

 Based on the above selection criteria, including the presence of H$\alpha$ and/or HI,  
we conducted 
an observational campaign to obtain optical spectra of the most promising TDG candidates.
A total of 11 candidate detached TDG were observed; of these, six were found to have redshifts
that matched the target galaxy pair, while five were found to have redshifts that did not match the target galaxy pair.
 
 The observations were carried out using the Boller \& Chivens spectrograph mounted on the 2.1m telescope at the San Pedro M\'artir Observatory. We utilized the 600l/mm grating  with a slit width of 3arcsec, providing a spectral resolution of approximately 
 $\sim4\rm{\AA}$ over a range spanning from 4800 to 7100 $\rm{\AA}$. A summary of the observations is presented in 
Table  \ref{tab_obs}. 
During each night of observations, a spectrophotometric standard star was observed to enable flux calibration 
of the 1D spectrum.
 The standard longslit spectral reduction techniques were applied to the data, including cosmic ray removal, bias subtraction, flat field correction, 
combination of multiple exposures, wavelength calibration, 1D spectra extraction, and sky subtraction.  IRAF was used for these tasks.

 \begin{table*}
 \caption{\label{tab_obs} Observing log and derived redshifts.}
{\centering
\begin{tabular}{cccccccc}
ID & RA & Dec &  Date & $t_{\rm{exp}}$ & $N_{\rm{obs}}$ & Spectral Range & $z$  \\
 & deg & deg & mm/dd/yyyy & min &  & $\rm{\AA}$ &  \\
\hline
Arp65a &5.4882353 & 22.3789987  & 10/30/2022 & 90 & 4  &  [4780-7100]  & 0.0185 $\pm$ 0.0003\tablefootmark{a}  \\
Arp72a & 236.7662451 & 17.8746735 & 05/04/2022 &  60 & 5 & [4780-7100] & 0.00113 $\pm$ 0.0003\tablefootmark{a} \\
Arp72b & 236.7195719 & 17.8894956  & 05/03/2022 & 60 & 5  & [4500-6820] & 0.0106 $\pm$ 0.0003\tablefootmark{a} \\
Arp72c & 236.7073667 & 17.8718278 & 05/05/2022 & 60 & 5  & [4780-7100] &  0.0111 $\pm$ 0.0003\tablefootmark{a}\\
Arp86a & 356.78143 & 29.448372  & 10/28/2022 &  60 & 3 & [4780-7100] &  0.0164 $\pm$ 0.0003\tablefootmark{a}\\
Arp86b &  356.76018 & 29.462161 & 10/29/2022 &  60 & 1 & [4780-7100] & 0.2482  $\pm$ 0.0004\tablefootmark{b}\\
Arp86d & 356.77929 & 29.467247 & 10/28/2022 &  60 & 3 & [4780-7100] &  0.0167 $\pm$ 0.0003\tablefootmark{a}\\
Arp86e & 356.74349 &29.439117 & 10/29/2022 &  60 & 1 & [4780-7100] & 0.1158 $\pm$ 0.0004\tablefootmark{b} \\
Arp87b & 175.17772 & 22.462017 & 5/04/2022 &  60 & 1 & [4780-7100] & 0.1119 $\pm$ 0.0004\tablefootmark{b} \\
Arp244a & 180.54481 & -18.910464 & 01/31/2022 &  60 & 2 & [4660-6980] & 0.1508 $\pm$ 0.0004\tablefootmark{b} \\
Arp256a & 4.6977042 & -10.36316 & 10/28/2022 &  45 & 1 & [4780-7100] & 0.1011 $\pm$ 0.0004\tablefootmark{b} \\

\end{tabular}
\tablefoot{
\tablefoottext{a}{Association with Parent Galaxies Confirmed by Redshift.}
\tablefoottext{b}{Confirmed Background Object.}
}}
\end{table*}  
 
 We found redshifts consistent with those of the possible host 
systems for six of the candidate TDGs
(Table  \ref{tab_obs}).  Three of these sources are associated with Arp 72, two with Arp 86, and one with
Arp 65. 
 Figure \ref{fig_obs} displays DESI-LIS color images of these three systems.
 The positions of the observed slits  are overlaid on the images.

\begin{figure*}
\centering
\includegraphics[width=0.9\linewidth]{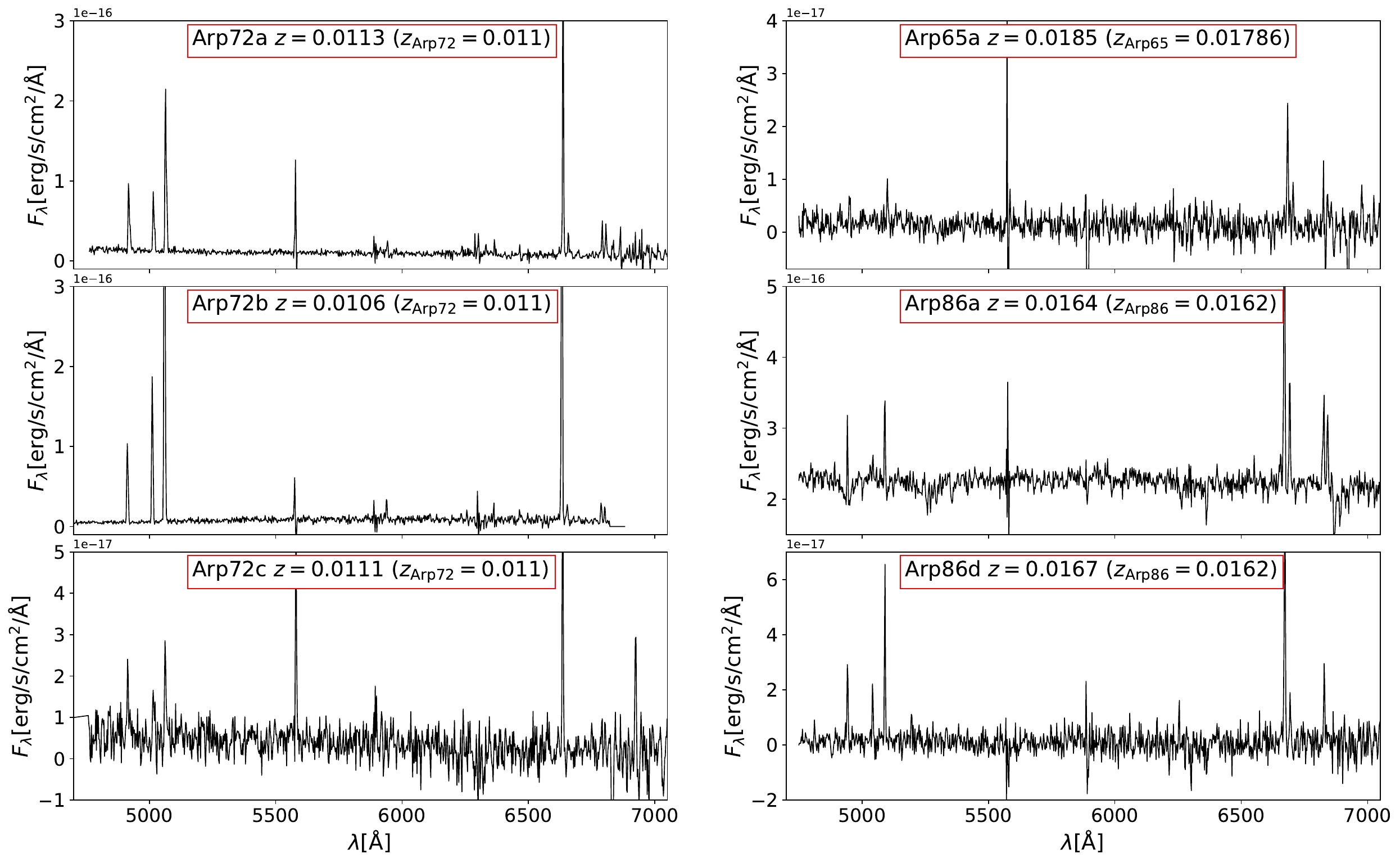}
\caption{\label{reduced_spectra} Spectra for the six TDGs 
confirmed to be at the redshift of the parent galaxies. 
In the red boxes, redshifts derived from H$\alpha$ are shown and compared with that of the host galaxy.}
\end{figure*}
 
 \begin{figure}
\centering
\includegraphics[width=0.9\linewidth]{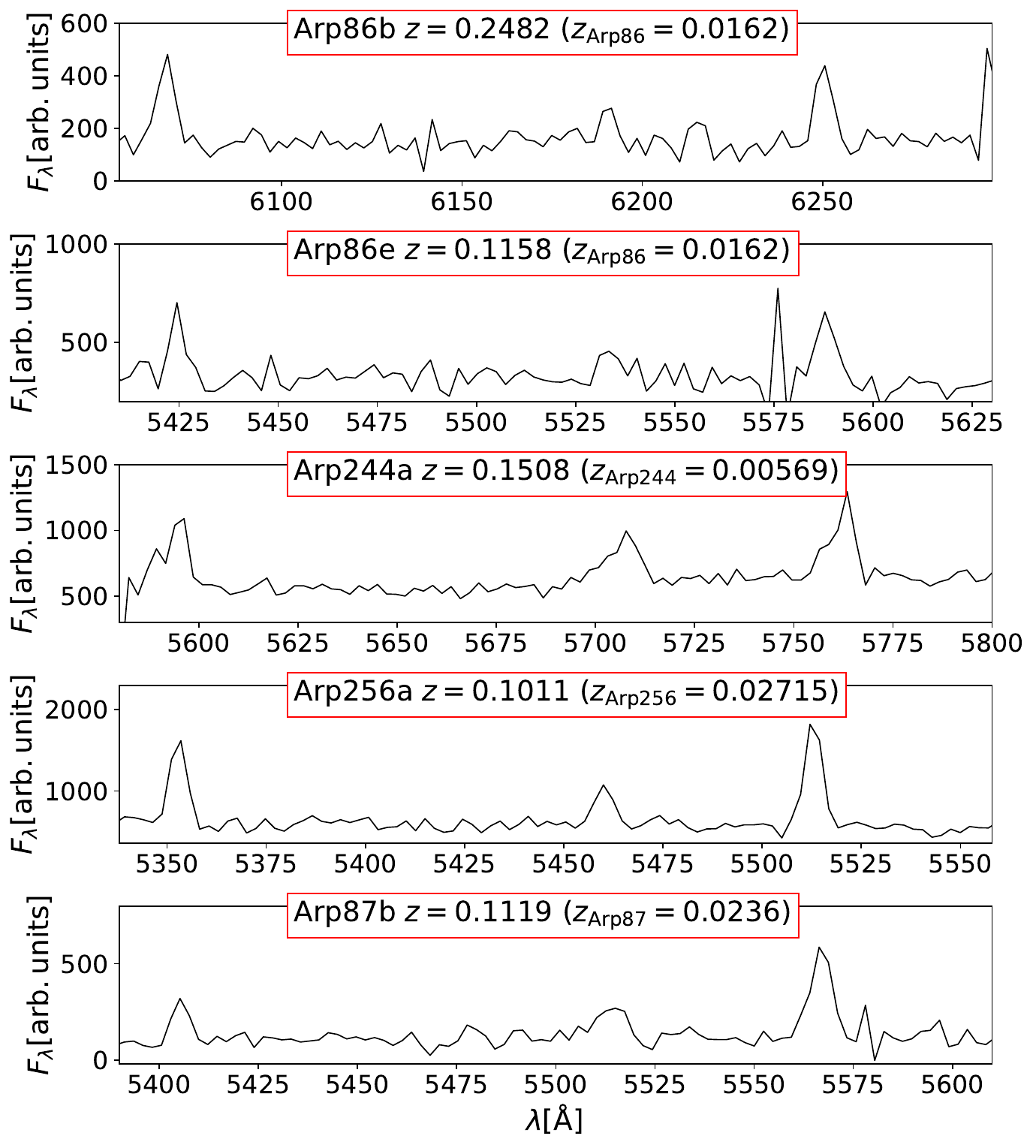}
\caption{\label{reduced_spectra_nontdgs} Spectra for the 5  objects
confirmed to not be at the redshift of the parent galaxies. 
In the red boxes, redshifts derived from H$\beta$ are shown and compared with that of the host galaxy.}
\end{figure}
 
 Figure \ref{reduced_spectra} showcases the reduced spectra of  these six objects. Emission lines are clearly detected in all objects. 
In this figure, we provide the redshift estimations based on the H$\alpha$ line and compare them with the values  for the parent galaxies reported in NASA NED. The observed redshifts indicate that these TDGs are associated with the interacting pair of galaxies, as the observed velocities are within a difference of 200 km/s or less.

On the other hand, we found that five of the detached TDG candidates are not at the redshift of their supposed parent galaxy (see Table \ref{tab_obs}). So, we confirm that these five objects are not detached TDGs, but background galaxies. We show the observed spectra of these five objects in Fig. \ref{reduced_spectra_nontdgs}, where the redshifted H$\beta$ and \oiii$\lambda$4959+$\lambda$5007 emission lines are clearly seen. The measured redshift for each TDG candidate is shown in Fig. \ref{reduced_spectra_nontdgs} together with the redshift of the supposed parent galaxy. The reason they have been observed just once ($N_{\rm{obs}}$) is because we could clearly see the emission lines immediately after the observation, so we decided to not continue the observation of these objects. The exception is Arp244a, since those observations were performed on service mode and we could not make decisions in real time. 

\begin{figure}
\centering
\includegraphics[width=0.88\linewidth]{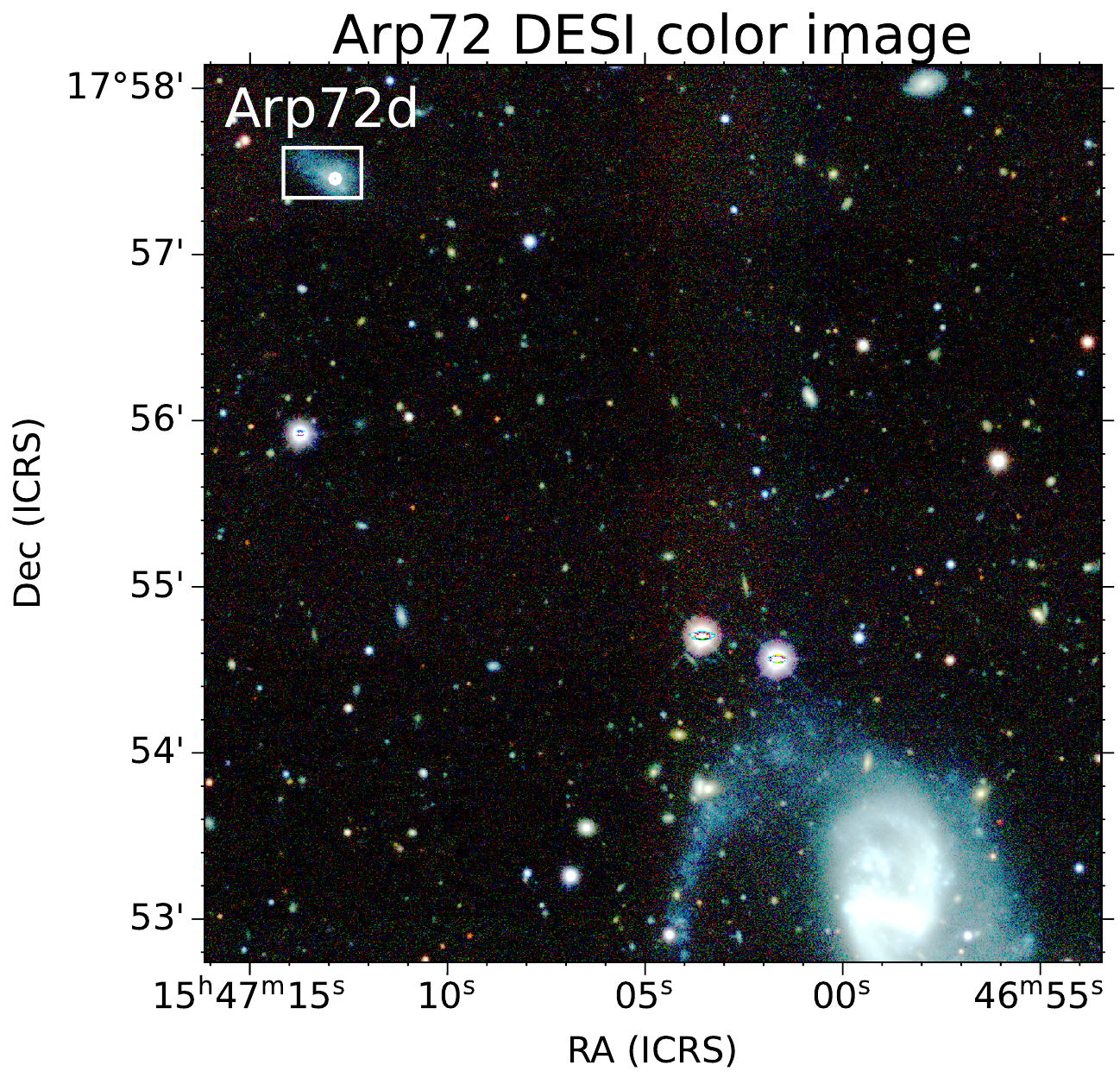} \\ 
\includegraphics[width=0.75\linewidth]{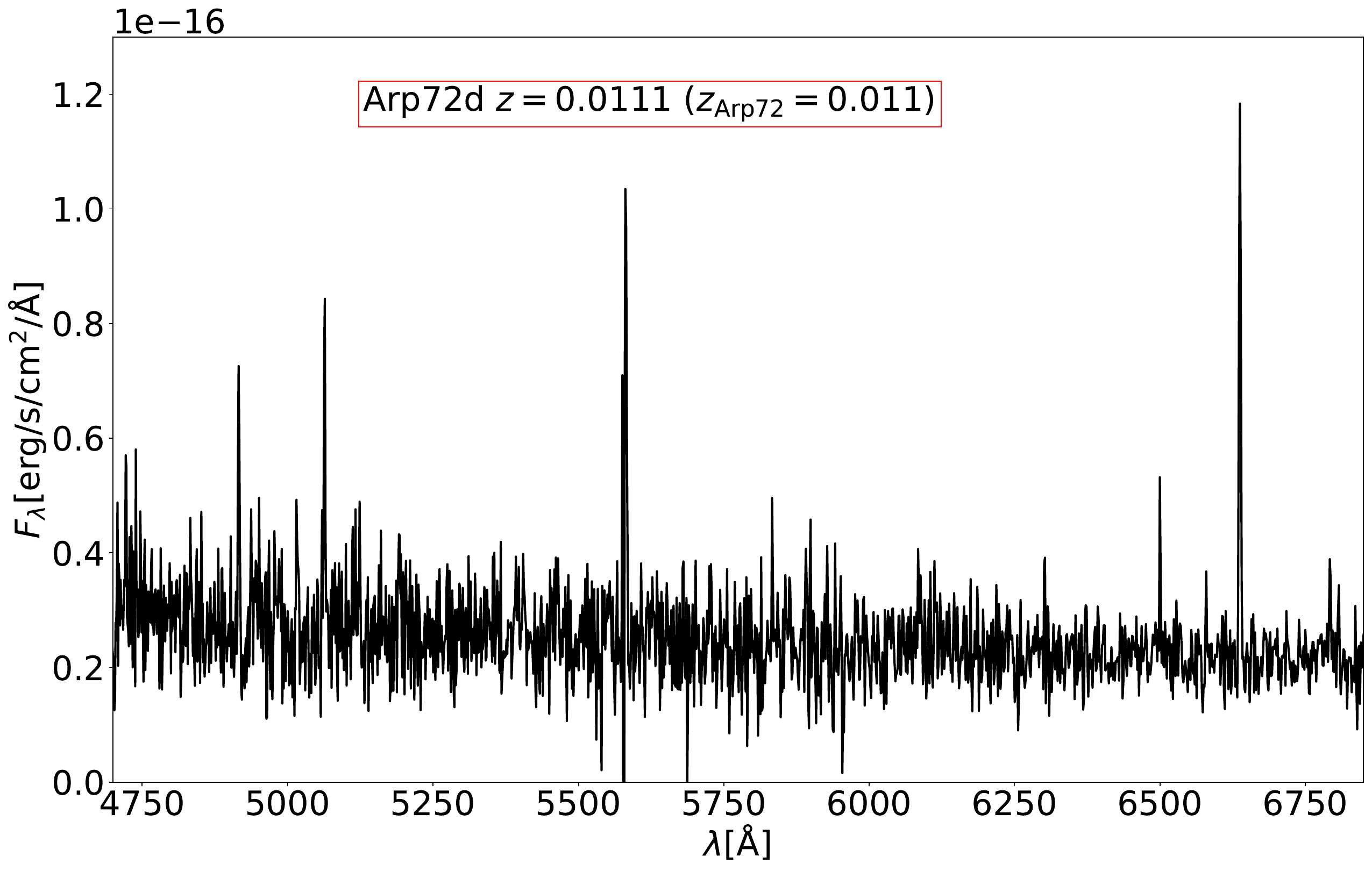} \\
\caption{\label{fig_arp72d} DESI-LIS color image of the Arp72d TDG candidate (top) and its optical spectrum from the SDSS (bottom). The TDG candidate is shown in the white square (northeast of Arp72), while the SDSS fiber where the spectrum is taken is shown as a white circle. }
\end{figure}



 Serendipitously, we found a detached TDG candidate near the Arp72 system in the DESI-LIS color images which is at the same redshift of Arp72. This source is hereafter identified as Arp72d. This TDG candidate was not in our list of candidates because it is not covered by \textit{Spitzer} 8 $\mu$m band  observations.
We show the Arp72d TDG candidate in Fig. \ref{fig_arp72d} together with the SDSS optical spectra of the object taken from SDSS-DR16 \citep{2020ApJS..249....3A}. The observed redshift  indicates that this TDG is also associated with the Arp72 system of interacting galaxies.

 It is possible that some of the confirmed detached TDGs may actually be preexisting dwarf galaxies  rather than TDGs. Metallicity  could be used as a potential discriminator in this regard. We have successfully detected multiple emission lines from these objects:
\hb, \ha,  \oiii$\lambda$4959+$\lambda$5007,  \nii$\lambda$6548+$\lambda$6583 and \sii$\lambda$6717+$\lambda$6731.  These
lines allowed us to estimate the oxygen abundance using 
 the  \cite{Pilyugin16}
strong line calibrator  method (Section 5.1).
 
The underlying stellar continuum is visible in Arp86a and to a lesser extent in Arp72a. In these two cases, we fitted the stellar continuum to the observed spectra 
using the Galaxy IFU Spectroscopy Tool (GIST) pipeline \citep{2019A&A...628A.117B}. The fitting procedure utilized penalised pixel-fitting (pPXF) code \citep{2017MNRAS.466..798C} and the Medium resolution INT Library of Empirical Spectra \citep[MILES,][]{2010MNRAS.404.1639V} to model the continuum and absorption lines. The fit was performed over the observed spectral range (refer to Table \ref{tab_obs}) using the spectral resolution measured as the median width of the lines observed in a lamp calibration spectrum, which yielded a  FWHM of $4.4\rm{\AA}$.

In this analysis, we focus on the emission lines and the removal of the stellar continuum, without examining the detailed results of the stellar continuum fitting. The primary objective is to recover the fluxes of the emission lines from the observed spectra. To estimate the fluxes from the pure emission line spectra, we employed Specutils,\footnote{\href{https://specutils.readthedocs.io}{https://specutils.readthedocs.io}} an associated package of Astropy \citep{2022ApJ...935..167A}. We utilized derivatives to locate local maxima, while using an emission-free spectral region to estimate the noise level. We considered peaks that were at least six times larger than the noise level as significant. Subsequently, we identified and measured the fluxes of the following lines: \hb, \ha,  \oiii$\lambda$4959+$\lambda$5007,  \nii$\lambda$6548+$\lambda$6583 and \sii$\lambda$6717+$\lambda$6731 (when detected). The integrated flux was estimated by integrating the flux over the relevant spectral range. To define this range, we identified the upper and lower wavelength limits by calculating the root mean square (rms) of the spectra in a nearby emission-free region and determined the first wavelengths from the maximum flux that crossed the rms. The spectral range used for estimating the integrated flux extended from the lower wavelength crossing the rms, subtracting 
$4.4\rm{\AA}$,
to the upper wavelength crossing the rms, adding 
$4.4\rm{\AA}$.
 The spectra were corrected for Galactic extinction before
line flux estimation.
We also corrected for internal dust extinction using the  H$\alpha$-H$\beta$ Balmer decrement, assuming $\frac{L_{\rm{H\alpha}}}{L_{\rm{H\beta}}}=2.863$ \citep{1989agna.book.....O} for case B recombination, and adopting attenuation curve values from \cite{2000ApJ...533..682C}, with $\kappa_{\rm{H\alpha}}=3.33$ and $\kappa_{\rm{H\beta}}=4.60$.

Table \ref{tab_fluxes} presents the measured extinction-corrected emission line fluxes for the confirmed detached TDGs, as well as for NGC 5994, the smaller companion of the Arp 72 system, where we utilized  the available spectra and emission line fluxes from SDSS centered in NGC5994.
 Table \ref{tab_fluxes} also provides E(B$-$V) for each source, as well as our derived values of log(O/H).
\begin{table*}
\fontsize{7.5}{10}\selectfont
 \caption{\label{tab_fluxes} Extinction corrected emission line fluxes for the confirmed detached TDGs and the Arp72 system smaller companion, NGC 5994. Fluxes are in units of $10^{-17}\rm{erg/s/cm^2}$. Color excess and oxygen abundance are also included.  }
\begin{tabular}{cccccccccccc}

 Name & H$\alpha$ & H$\beta$ & [OIII]4959 & [OIII]5007 & [NII]6548 & [NII]6583 & [SII]6717 & [SII]6731 & E(B-V) & 12+log(O/H)\\

\hline
Arp65a & $14.6 \pm 1.2$ & $5.4 \pm 1.4$ & $ ... $ & $6.4 \pm 1.3$ & $ ... $ & $5.9 \pm 1.2$ & $4.8 \pm 1.2$ & $4.3 \pm 1.4$ & $ ... $ & $8.5\pm0.7$ \\
Arp72a & $176.4 \pm 0.8$ & $63.2 \pm 1.1$ & $47.7 \pm 1.1$ & $132.2 \pm 1.0$ & $4.6 \pm 0.8$ & $13.8 \pm 0.8$ & $20.6 \pm 0.9$ & $19.8 \pm 0.9$ & $ ... $ & $8.14\pm0.1$ \\
Arp72b & $1693.5 \pm 2.0$ & $591.4 \pm 2.3$ & $977.3 \pm 2.2$ & $3132.7 \pm 2.2$ & $ ... $ & $63.4 \pm 2.3$ & $59.7 \pm 2.1$ & $36.5 \pm 1.6$ & $0.524 \pm 0.009$ & $8.14\pm0.16$ \\
Arp72c & $35.1 \pm 1.6$ & $19.0 \pm 2.2$ & $13.7 \pm 2.2$ & $19.9 \pm 2.0$ & $ ... $ & $ ... $ & $ ... $ & $ ... $ & $ ... $ & $...$ \\
Arp72d & $42.3 \pm 1.6$ & $18.4 \pm 1.8$ & $5.62 \pm 0.09$ & $17.02 \pm 0.09$ & $ ... $ & $2.3 \pm 0.4$ & $6.15 \pm 0.16$ & $4.47 \pm 0.22$ & $ ... $& $7.5\pm1.0$ \\
NGC5994 & $2831 \pm 4$ & $989 \pm 5$ & $769 \pm 5$ & $2241 \pm 5$ & $55.6 \pm 3.2$ & $316 \pm 4$ & $605 \pm 4$ & $463 \pm 5$ & $0.372 \pm 0.007$ & $8.21\pm0.05$\\
Arp86a & $620 \pm 4$ & $217 \pm 4$ & $46.5 \pm 3.4$ & $173 \pm 4$ & $82 \pm 5$ & $184 \pm 4$ & $158 \pm 4$ & $110 \pm 4$ & $0.233 \pm 0.020$ & $8.43\pm0.07$ \\
Arp86d & $73.5 \pm 1.9$ & $25.7 \pm 1.7$ & $18.1 \pm 1.7$ & $45.9 \pm 1.6$ & $7.5 \pm 2.2$ & $14.3 \pm 1.9$ & $21.7 \pm 1.8$ & $4.1 \pm 1.2$ & $0.11 \pm 0.05$ & $8.3\pm0.4$\\
\end{tabular}
\end{table*}

 To compare the emission line properties of the TDGs with those of the host galaxies (section 6.2), it is
necessary for us to obtain optical line fluxes and derive metallicities 
for star-forming regions within the main disks of the
galaxies as well.
For Arp 72, we combined the observed spectra with  archival data for the main galaxy, NGC 5995, obtained by the Multi Unit Spectroscopic Explorer  \citep[MUSE, ][]{2010SPIE.7735E..08B}. MUSE is an integral field spectrograph installed on the 
Very Large Telescope (VLT), offering a field of view of $1\times1~\rm{arcmin}^2$, a spectral range between 4800 and 9300 $\rm{\AA}$, and a spectral resolution ranging from 2.4 to 3 $\rm{\AA}$. The MUSE observation of Arp 72 covers the white square area shown in Figure \ref{fig_obs} (middle). We performed stellar continuum subtraction as described earlier, while considering the different spectral resolution of the MUSE data. The stellar continuum was fitted between 4800 and 6850 $\rm{\AA}$, aligning the minimum spectral resolution of 2.5 Å with that of the MILES library. Subsequently, we estimated the fluxes for the same emission lines as the TDGs, following the same procedure. For Arp 86, we combined our flux measurements with those 
reported in \cite{2014RAA....14.1393Z}.  For Arp 65, we used the reported oxygen abundances from  \cite{2014RAA....14.1393Z} and \cite{2020MNRAS.498..101Z}, as they employed the same calibrator we used, the S-calibrator from \cite{Pilyugin16}.

\begin{table*}
 \fontsize{7.5}{10}\selectfont
 \caption{\label{tab_phot}GALEX, {\it \textit{Spitzer}}, and DESI-LIS photometry of the spectroscopically confirmed detached TDGs.}
 \begin{tabular}{cccccccc}
 id & FUV & NUV & u (SDSS) & g (SDSS) & r (SDSS) & i (SDSS) & z (SDSS) \\
 & mJy & mJy & mJy & mJy & mJy & mJy & mJy  \\
\hline
Arp65a & $0.00989 \pm 0.00026$ & $0.00863 \pm 0.00032$ & $< 0.03$ & $0.0119 \pm 0.0021$ & $< 0.01$ & $< 0.03$ & $< 0.08$ \\
Arp72a & $0.0353 \pm 0.0005$ & $0.0323 \pm 0.0004$ & $0.042 \pm 0.008$ & $0.0623 \pm 0.0027$ & $0.082 \pm 0.004$ & $0.069 \pm 0.007$ & $< 0.1$ \\
Arp72b & $0.01165 \pm 0.00032$ & $0.01105 \pm 0.00031$ & $< 0.06$ & $0.0713 \pm 0.0025$ & $0.063 \pm 0.004$ & $0.067 \pm 0.007$ & $< 0.1$ \\
Arp72c & $0.01067 \pm 0.00031$ & $0.00934 \pm 0.00033$ & $< 0.03$ & $0.0139 \pm 0.0019$ & $0.0106 \pm 0.0031$ & $< 0.03$ & $< 0.07$ \\
Arp72d & $0.04851 \pm 0.00033$ & $0.05901 \pm 0.00035$ & $0.167 \pm 0.008$ & $0.2948 \pm 0.0024$ & $0.354 \pm 0.004$ & $0.387 \pm 0.006$ & $0.391 \pm 0.033$ \\
Arp86a & $0.03722 \pm 0.00032$ & $0.0428 \pm 0.0005$ & $0.190 \pm 0.004$ & $0.6289 \pm 0.0016$ & $1.1099 \pm 0.0030$ & $1.422 \pm 0.005$ & $1.756 \pm 0.019$ \\
Arp86d & $0.0128 \pm 0.0004$ & $0.0125 \pm 0.0005$ & $< 0.03$ & $0.0128 \pm 0.0017$ & $< 0.04$ & $< 0.06$ & $< 0.2$ \\
 \end{tabular}
\end{table*}


\begin{table*}
 \fontsize{7.5}{10}\selectfont
 \caption*{Table \ref{tab_phot}, continued.}
 \begin{tabular}{ccccccccc}

 id & g (DES) & r (DES) & z (DES) & $3.6\rm{\mu m}$ & $4.5\rm{\mu m}$ & $5.8\rm{\mu m}$ & $8.0\rm{\mu m}$ & $24\rm{\mu m}$ \\
 & mJy & mJy & mJy & mJy & mJy & mJy & mJy & mJy   \\
\hline
Arp65a & $0.0086 \pm 0.0005$ & $0.0095 \pm 0.0008$ & $0.0094 \pm 0.0022$ & $0.0321 \pm 0.0018$ & $0.0197 \pm 0.0022$ & $0.046 \pm 0.010$ & $0.113 \pm 0.008$ & $< 0.2$ \\
Arp72a & $0.0754 \pm 0.0008$ & $0.0931 \pm 0.0015$ & $0.108 \pm 0.004$ & $0.0408 \pm 0.0019$ & $0.0208 \pm 0.0022$ & $< 0.02$ & $0.039 \pm 0.009$ & $< 0.1$ \\
Arp72b & $0.0508 \pm 0.0007$ & $0.0693 \pm 0.0014$ & $0.081 \pm 0.004$ & $0.0255 \pm 0.0016$ & $0.0174 \pm 0.0020$ & $0.065 \pm 0.008$ & $0.053 \pm 0.008$ & $0.23 \pm 0.04$ \\
Arp72c & $0.0194 \pm 0.0011$ & $0.0222 \pm 0.0021$ & $< 0.02$ & $0.0166 \pm 0.0015$ & $< 0.006$ & $< 0.02$ & $0.038 \pm 0.008$ & $< 0.1$ \\
Arp72d & $0.2827 \pm 0.0007$ & $0.3652 \pm 0.0014$ & $0.408 \pm 0.004$ & $ ... $ & $ ... $ & $ ... $ & $ ... $ & $ ... $ \\
Arp86a & $0.6608 \pm 0.0005$ & $1.1842 \pm 0.0014$ & $1.7109 \pm 0.0025$ & $0.9117 \pm 0.0019$ & $0.5463 \pm 0.0024$ & $0.684 \pm 0.011$ & $1.200 \pm 0.011$ & $0.96 \pm 0.07$ \\
Arp86d & $0.0160 \pm 0.0005$ & $0.0182 \pm 0.0012$ & $< 0.008$ & $0.0241 \pm 0.0020$ & $0.0123 \pm 0.0025$ & $0.081 \pm 0.012$ & $0.110 \pm 0.014$ & $< 0.2$ \\
 \end{tabular}
\end{table*}

We  provide the UV/optical/IR photometry of  the final sample 
of redshift-confirmed detached TDGs in Table \ref{tab_phot}.
 The fluxes in Table \ref{tab_phot} have been corrected for
Galactic extinction.
 In Section 5.2, 
we use this photometry to compute and fit the SEDs of the TDGs.

\section{Detailed description of systems}

\subsection{Arp 65}
Arp 65 (Figure \ref{fig_obs}, top panel) consists of two spirals, NGC 90 to the west and NGC 93 to the east.
NGC 93 shows two long straight tidal tails extending from a "grand design" two-armed spiral pattern in the 
inner disk.
The edge-on disk galaxy NGC 93 has a 3.5 $\mu$m flux three times that of the western galaxy \citep{smith2007}
and an estimated stellar mass three times larger \citep{2015A&A...584A.114S}. 
A 21 cm HI map of the pair has been presented by \cite{2015A&A...584A.114S},  
who conclude that NGC 93 is gas-poor and NGC 90 gas-rich.  Both have peculiar HI morphologies.  
The HI associated with NGC 93 is skewed to the north relative to the optical galaxy,
while a massive concentration of HI (3.4 $\pm$ 0.4 $\times$ 10$^9$ M$_{\sun}$) is offset to the southeast of 
NGC 90.
\cite{2015A&A...584A.114S} conclude that the most likely explanation for the large concentration of gas
outside of NGC 90 is that it was tidally stripped from NGC 90 
about 100 $-$ 250 Myrs ago
by the interaction with NGC 93.
Alternatively, 
\cite{2020MNRAS.498..101Z} 
suggest that this gas was removed by ram pressure stripping due to motion through intragroup gas.

Extra-nuclear knots of star formation are detected near the base of the tidal features of NGC 90
in the UV
\citep{smith2010,
2015A&A...584A.114S} and in optical emission lines including H$\alpha$ 
\citep{2020MNRAS.498..101Z}.  
Our detached TDG lies further away from NGC 90 than these sources, near the southern end of the HI 
concentration.  Our best-fit determination of the starburst age of this TDG (see Section \ref{sec:SED}), 110 $\pm$ 50 Myrs, is consistent with
the 
\cite{2015A&A...584A.114S} 
interaction age, supporting the idea that this star formation was triggered by the interaction. 

\subsection{Arp 72}

The M51-like galaxy Arp 72  (Fig. \ref{fig_obs}, middle panel)  is composed of the larger galaxy NGC 5996 in the east and a small companion
NGC 5994 
to the west,
connected by an optical bridge.
The 3.6 $\mu$m flux of the primary is 12 times that of the companion \citep{smith2007}.
In UV images, a long (3$'$; 47 kpc) curved tidal tail 
extends to the east from the primary galaxy \citep{smith2010}; this tail is also visible in optical pictures
but is not as prominent in the optical as in the UV.  
The 21 cm HI map of \cite{2012MNRAS.420....2S} shows that this tail is also rich in HI.
Another HI tail
extends 1$'$ (15 kpc) to the west of the companion \citep{2012MNRAS.420....2S}.

TDG Arp72a is located within the long HI-rich tail to the east, and is clearly tidal in origin.
However, the optical surface brightness of the surrounding tail is below our nominal cut-off, and thus
by our definition, this TDG is classified as "detached."   
TDG Arp72b is located near the end of a shorter extension of HI to the northeast,
while TDG Arp72c is within the western HI tail.
 Arp72b and Arp72c are clearly tidal in origin, and are unambiguously "detached" in the sense that
they are not connected to the main galaxy by an optically visible tail.  However, they are still connected
to the main galaxy via gaseous features. Arp72d is located at 5$'$ (80 kpc) from the main galaxy, NGC5996.  Arp72d is not in the field of view of the HI map presented in \citet{2012MNRAS.420....2S}. Within the field of view, no HI tail connects the main galaxy to the area where Arp72d is located. 

Near the position of Arp 72c, the HI velocity field of 
\cite{2012MNRAS.420....2S} at 40$''$ resolution shows a gradient, with
positive velocities relative to the systematic velocity 
to the east of Arp 72c and negative velocities to the west.  Over
a scale of about 20$''$ east-west, 
the radial velocity changes by about 20~km~s$^{-1}$.  However,
this trend may be part of a smooth gradient extending from
the eastern side of 
the companion galaxy NGC 5994, rather than a signature of
rotation within the TDG.
Higher spatial resolution data is needed to test this.

\subsection{Arp 86}

Arp 86 (Fig. \ref{fig_obs}, bottom panel) is another M51-like interacting pair. The primary galaxy NGC 7753 in the north has a 3.6 $\mu$m
flux five times that of the smaller companion NGC 7752 in the south. The HI map of
\cite{2009MNRAS.397..548S}
reveals a HI countertail to the north, a tail that is not
present in UV or optical maps \citep{smith2010}.
To the southeast of the companion, 
two parallel HI tails
are seen
\citep{2009MNRAS.397..548S}.
From H$\alpha$ rotation curves, \cite{marcelin1987} infer
that the ratio of the dynamical masses of the two galaxies
in the pair is 1:10.

The candidate TDG we name Arp86a is listed in the 2MASS
catalog as 2MASX J23470758+2926531.  This object has 
previously been
studied by 
\cite{2014RAA....14.1393Z}, who find a redshift
and metallicity consistent with our values. 
\cite{2009MNRAS.397..548S} find an HI counterpart to this source,
and derive an HI mass of 4.5 $\times$ 10$^8$~M$_{\sun}$.
They state that the HI spectrum resembles a weak double horn
consistent with rotation, but we note that its spectrum might be
affected by tidal features near the galaxy.
As noted below (Section \ref{sec:mass_metallicity}), in metallicity-M* relations, this object
lies in the region populated by normal dwarf galaxies; thus, it is likely
a preexisting dwarf rather than a true TDG.

Confirmed TDG Arp86d is a relatively low stellar mass 
(3.6 $\pm$ 1.2 $\times$ 10$^6~{M_{\sun}}$) object 
 (see Section \ref{sec:SED}) just outside 
of the main disk of NGC 7753.  It appears to be part of a faint
outer spiral arm of this galaxy seen in UV maps
\citep{smith2010}.

We also obtained optical spectra of two other candidate TDGs
in the vicinity of Arp 86, but both were found to be
background objects (see Table \ref{tab_obs}). 

\section{Analysis}
\subsection{Oxygen nebular abundance}

 The discrimination between TDGs and ordinary dwarf galaxies
using the nebular  oxygen-to-hydrogen (O/H) abundance-stellar mass relation depends on reliably distinguishing 
low and high oxygen abundances. We used  the 
\cite{Pilyugin16}
strong line calibrator, which is capable of doing so. 
Since the [OIII]4959 and [NII]6548 lines are marginally detected  from
our sources (if detected), we used the theoretical relation of one-third with respect to their doublet-companion lines, [OIII]5007 and [NII]6583, respectively \citep{2017PASP..129d3001P}.
The derived oxygen abundances are given in Table \ref{tab_fluxes}. 
We used the same method  to obtain abundances
for the star-forming regions within the  main galaxies of Arp 65 and Arp 72.

\subsection{Ultraviolet to mid-infrared spectral energy distribution fitting} \label{sec:SED}

We have utilized our photometry to perform stellar  population and dust emission fitting using the Code Investigating GALaxy Emission \citep[CIGALE,][]{2019A&A...622A.103B}. The observational data used in the fitting process are listed in Table \ref{tab_phot}.

\begin{figure*}
\centering
\begin{tabular}{cc}
\includegraphics[width=0.43\linewidth]{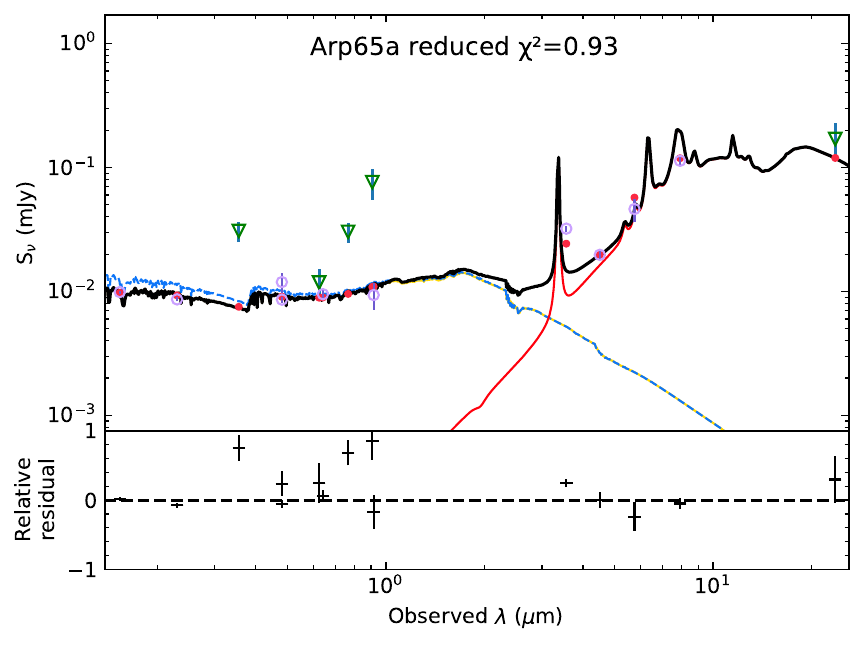}&
\includegraphics[width=0.3\linewidth]{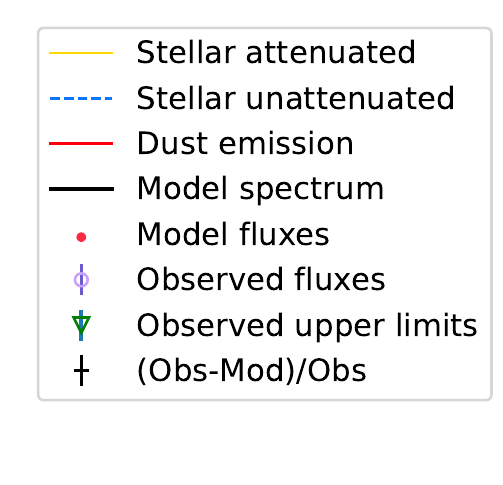}\\
\includegraphics[width=0.43\linewidth]{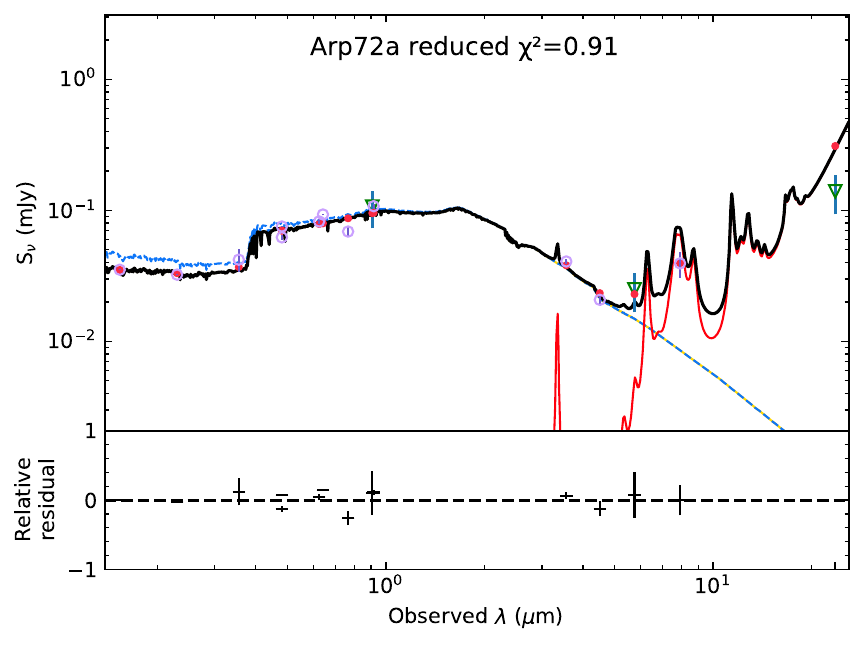}&
\includegraphics[width=0.43\linewidth]{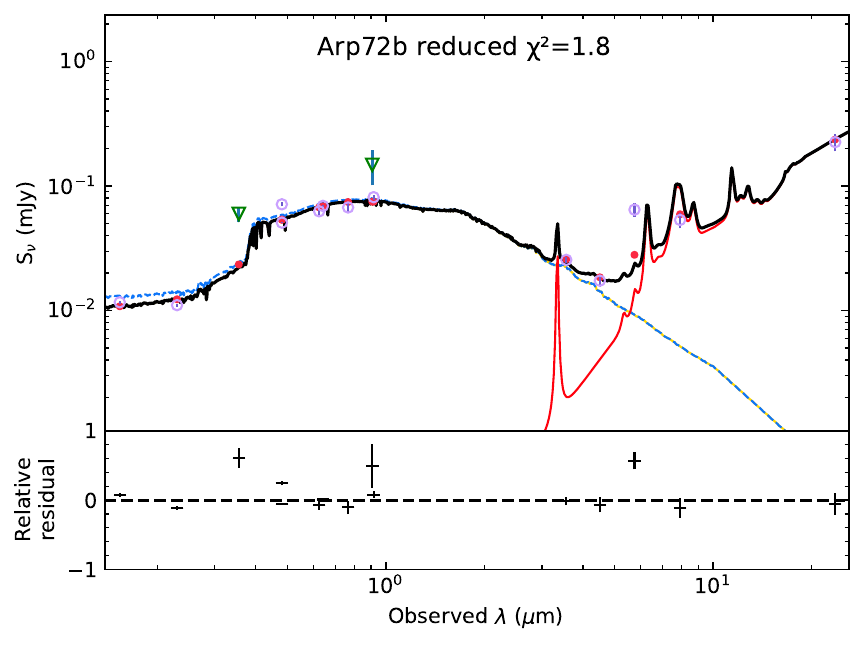}\\
\includegraphics[width=0.43\linewidth]{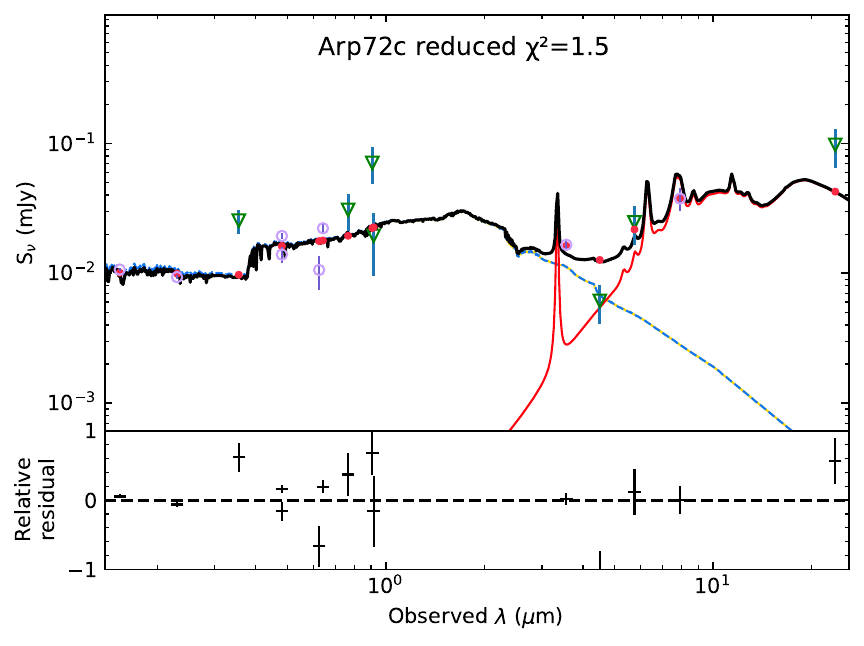}&
\includegraphics[width=0.43\linewidth]{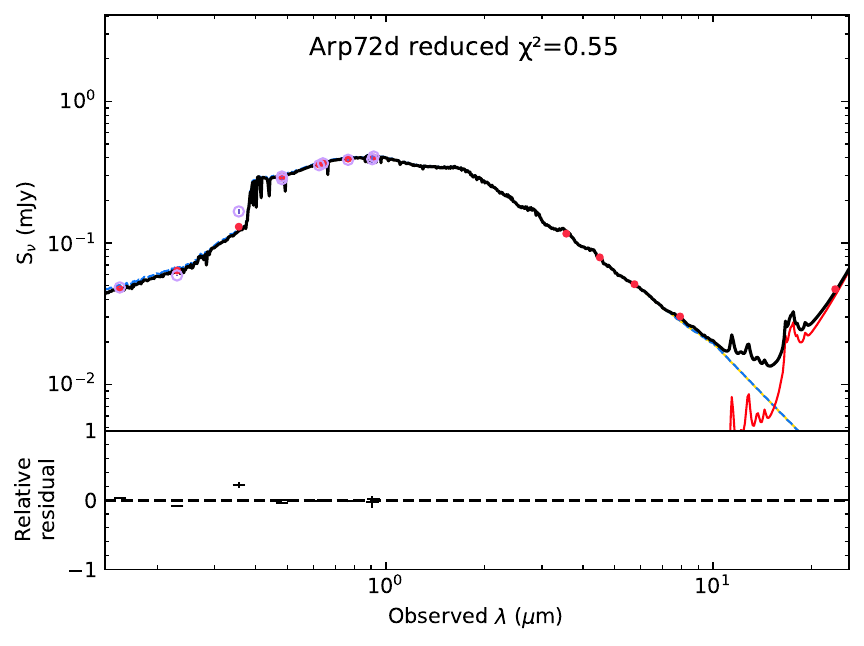}\\
\includegraphics[width=0.43\linewidth]{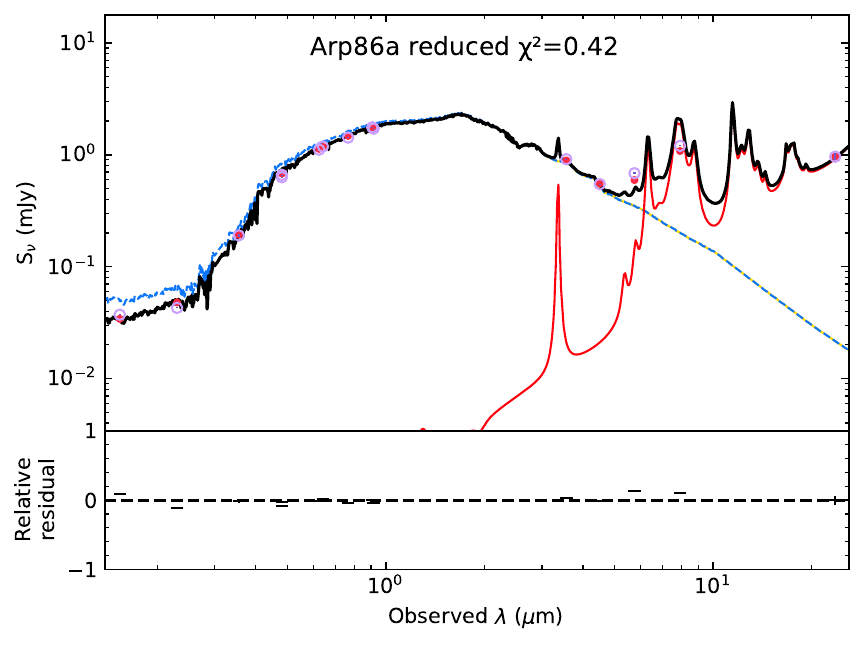}&
\includegraphics[width=0.43\linewidth]{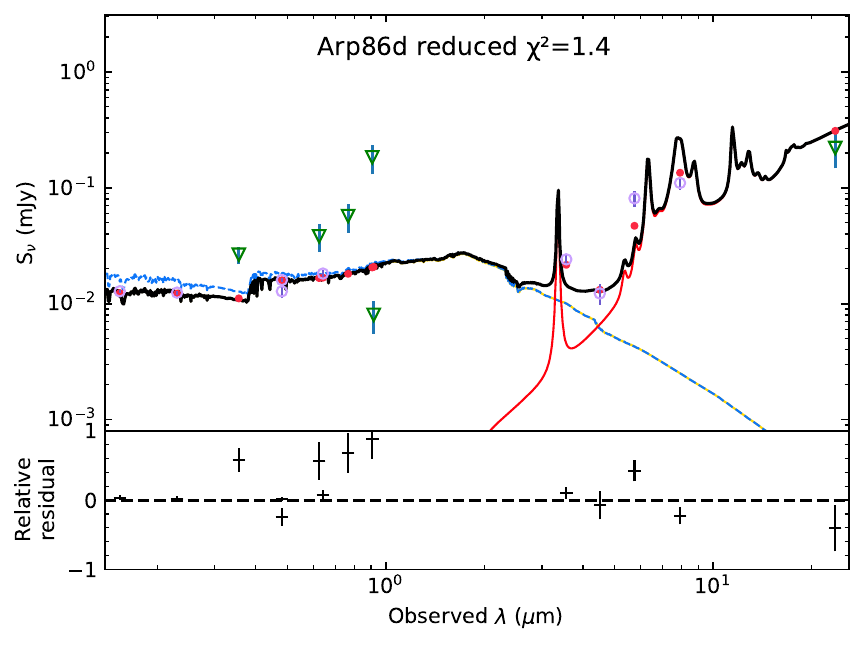}\\
\end{tabular}
\caption{\label{fig_seds} Spectral energy distributions of the seven spectroscopically confirmed TDGs. We show the fitted model using CIGALE as black lines. }

\end{figure*}

For the modeling, we assumed a double exponential star formation history, a modified dust law based on \cite{2000ApJ...533..682C} and \cite{2002ApJS..140..303L}, single stellar populations (SSPs) from \cite{2003MNRAS.344.1000B}, and the dust emission model proposed by \cite{2014ApJ...780..172D}. The allowed parameters for the different models are presented in Table \ref{tab_inputcig}.
The results obtained from the CIGALE fits are summarized in Table \ref{tab_outputcig}, and the observed and fitted SEDs are shown in Figure \ref{fig_seds}.

\begin{table*}
{\scriptsize
\caption{CIGALE input parameters\label{tab_inputcig}.}
 \centering
 \begin{tabularx}{\textwidth}{lX}
 \hline
  Free parameters&\\\hline\hline
  $e$-folding time of the old population& 1, 2, 3, 4, 5 Gyr\\
  $e$-folding time of the late starburst population & 25, 50, 100, 125, 150, 175, 200, 250, 300, 400, 500 Myr\\
  Mass fraction of the late burst population &  0.1, 0.2, 0.3, 0.4, 0.5, 0.6, 0.7, 0.8, 0.9\\
  Age of the late burst& 25, 50, 100, 150, 200, 300, 400, 500, 600, 700, \\& 800, 900, 1000, 1250, 1500, 2000 Myr\\
  Metallicity& 0.0001, 0.0004, 0.004, 0.008, 0.02, 0.05\\
  E(B$-$V) of the stellar continuum light for the young population.& 0.01, 0.02, 0.03, 0.04, 0.05,  0.06 mag\\
  
  Mass fraction of PAH & 0.47, 1.12, 3.19, 5.26, 7.32\\
  Powerlaw slope ${\rm{d}}U/{\rm{d}}M\propto U^{\alpha}$ &  1, 2, 3\\
  Fraction illuminated from minimum to maximum radiation field & 0.0, 0.25, 0.5, 0.75, 1.0\\
  Minimum radiation field &  0.1, 0.4, 0.8, 1, 2, 4, 8, 12, 17, 20, 25, 30, 50\\
  
\hline
  Fixed parameters&\\\hline\hline
  Age of the oldest stars&11 Gyr\\
  Reduction factor for the E(B-V) of the old population compared to the young one &0.5\\
  Central wavelength of the UV bump& 217.5 nm \\ 
   Amplitude of the UV bump& 35 nm \\ 
  Slope $\delta$ of the power law modifying the attenuation curve&0\\ 
    IMF & \cite{2003PASP..115..763C}\\

\hline
 \end{tabularx}
 }
\end{table*}

\begin{table*}

\caption{ \label{tab_outputcig} Results from CIGALE fits.}
\begin{tabular}{cccccc}
id & $M_{*}$ & $Z$ & SFR & $f_{\rm{burst}}$\tablefootmark{a} & burst age \\
 & $\rm{M_{\odot}}$ &  &$\rm{M_{\odot}}/\rm{yr}$ & & Myr\\ 
 \hline
Arp65a & $\left(1.7 \pm 1.1\right) \times 10^{6}$ & $0.023 \pm 0.019$ & $0.0060 \pm 0.0009$ & $0.47 \pm 0.26$ & $\left(1.1 \pm 0.5\right) \times 10^{2}$ \\
Arp72a & $\left(1.2 \pm 0.4\right) \times 10^{7}$ & $0.007 \pm 0.005$ & $0.0060 \pm 0.0013$ & $0.41 \pm 0.25$ & $\left(4.3 \pm 1.5\right) \times 10^{2}$ \\
Arp72b & $\left(1.7 \pm 0.5\right) \times 10^{7}$ & $0.0005 \pm 0.0009$ & $0.0010 \pm 0.0004$ & $0.36 \pm 0.23$ & $\left(8.7 \pm 3.3\right) \times 10^{2}$ \\
Arp72c & $\left(2.5 \pm 1.3\right) \times 10^{6}$ & $0.034 \pm 0.018$ & $0.0021 \pm 0.0004$ & $0.45 \pm 0.26$ & $\left(3.2 \pm 1.1\right) \times 10^{2}$ \\
Arp72d & $\left(8.4 \pm 2.7\right) \times 10^{7}$ & $0.001 \pm 0.004$ & $0.0034 \pm 0.0020$ & $0.46 \pm 0.24$ & $\left(8.7 \pm 3.3\right) \times 10^{2}$ \\
Arp86a & $\left(1.09 \pm 0.20\right) \times 10^{9}$ & $0.0060 \pm 0.0025$ & $0.0173 \pm 0.0032$ & $0.29 \pm 0.17$ & $\left(1.85 \pm 0.27\right) \times 10^{3}$ \\
Arp86d & $\left(3.6 \pm 2.2\right) \times 10^{6}$ & $0.023 \pm 0.019$ & $0.0068 \pm 0.0015$ & $0.42 \pm 0.27$ & $\left(1.7 \pm 0.6\right) \times 10^{2}$ \\
\end{tabular}
\tablefoot{
\tablefoottext{a}{Fraction of stars formed in the recent burst relative to the total mass of stars ever formed.}
}
\end{table*}

\begin{figure}
\centering
\includegraphics[width=0.9\linewidth]{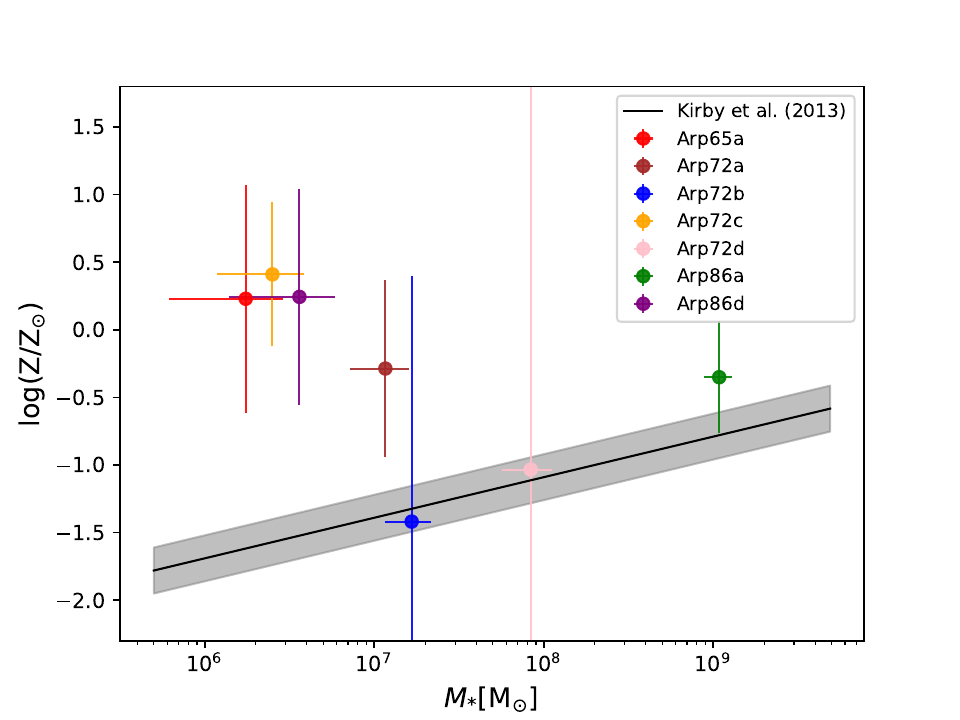}
\caption{\label{fig_metmass} 
Stellar metallicity versus stellar mass for the six redshift confirmed TDGs. The fundamental relation for dwarf galaxies obtained from \cite{2013ApJ...779..102K} is shown as a black line. 
 The gray shaded range marks the rms spread in the best-fit relation.
  The metallicity of the Sun is set to $Z_{\odot} = 0.134$ \citep{2009ARA&A..47..481A}.
}
\end{figure}

\section{Results}

\subsection{Comparison with standard mass-metallicity relations for dwarf galaxies} \label{sec:mass_metallicity}

The fundamental relation between metallicity and stellar mass 
 could be used as a discriminator between TDGs and preexisting dwarf galaxies. In Figure \ref{fig_metmass}, we plot the stellar metallicity versus stellar mass for the TDGs. We observe a clear division in stellar mass, where a group of six TDGs has stellar masses ($M_{*}$) less than $1 \times 10^8~\rm{M_{\odot}}$, and one TDG (Arp86a) stands out as significantly more massive with $M_{*}=1.4  \times 10^9~\rm{M_{\odot}}$. Comparing these results with the relation for dwarf galaxies from \cite{2013ApJ...779..102K},  we see that four sources, Arp 65a, Arp 72c, Arp86d, and Arp72a 
lie significantly above the standard relation, taking into account the uncertainties in the stellar metallicities obtained from CIGALE. 
For the remaining TDGs, however, there are large enough uncertainties in the
metallicity to make it difficult to discriminate between the two options. 
It is important to note that the relation from \cite{2013ApJ...779..102K} uses iron abundances, [Fe/H], rather than direct metallicity values $Z$. However, for a first-order approximation, these values can be considered equivalent.

\begin{figure}
\centering
\includegraphics[width=0.9\linewidth]{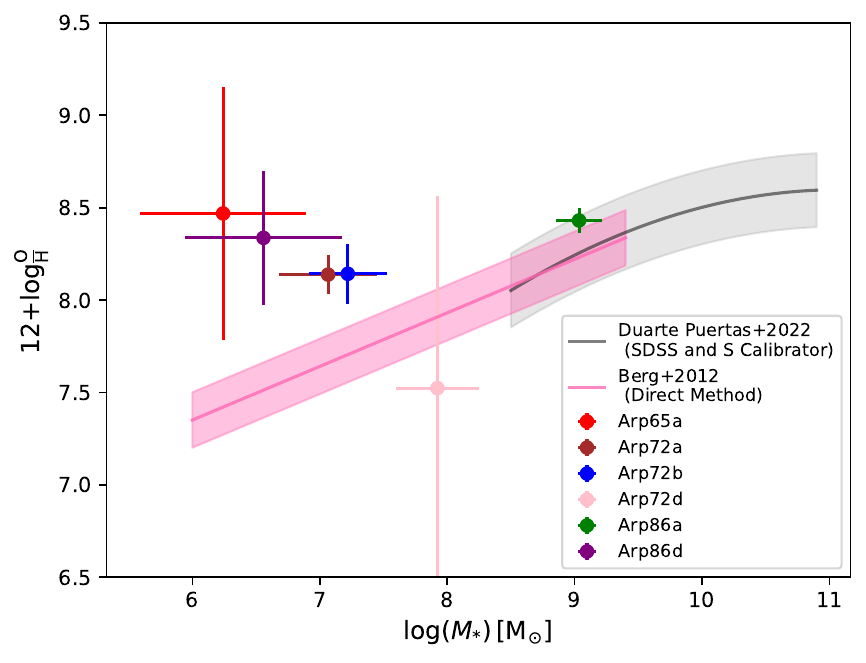}
\caption{\label{fig_oabundance} Nebular oxygen abundance versus stellar mass for the six spectroscopically confirmed TDGs. No oxygen abundance is available for Arp72c.
The fundamental relation for  star-forming
SDSS galaxies  with log M* $>$ 8.5
from \cite{2022A&A...666A.186D} is shown as a black  solid line.  
 The gray shaded range marks the rms spread in the best-fit relation.
 Dwarf galaxy relation from \cite{berg2012}
is shown in pink.
}
\end{figure}

The nebular oxygen abundance versus stellar mass is plotted in Fig. \ref{fig_oabundance}. 
 We overlap on this plot 
the relations reported by \cite{2022A&A...666A.186D} for star-forming SDSS galaxies with
$\rm{log} (M_{*}) > 8.5$, which was 
also derived using the 
\cite{Pilyugin16} strong-line method.
We have also overlayed the \cite{berg2012} 
relation for dwarf galaxies. The latter curve was obtained using
oxygen abundances derived 
using the "direct" method of deriving oxygen abundances 
(i.e., using electron temperatures calculated from the
\oiii$\lambda~5007/$\oiii$\lambda~4363$ emission line ratio).
According to 
\cite{Pilyugin16}, oxygen abundances from their strong-line method agrees with 
the "direct" method within 0.1 dex.
In 
Fig. \ref{fig_oabundance},
the group of less massive TDGs clearly exhibits higher oxygen abundances compared to  other dwarf galaxies with similar masses, even when accounting for the additional uncertainty of 0.1 dex due to the different methods of obtaining the oxygen abundances.
However, the most massive object, Arp86a, might not be a true TDG since its oxygen abundance is quite close to that expected in the fundamental relation.   Arp72d has a low oxygen abundance, even for its mass, although the uncertainty in the oxygen abundance value is very high and it could still be a detached TDG. Arp72c oxygen abundance is not possible to determine due to the lack of [NII]6583 line detection, but since the stellar metallicity is large, we conclude that is in fact a detached TDG.

In conclusion, there is evidence from the Mass-Metallicity and Mass-Oxygen abundance relations that the five least massive objects are detached TDGs since either the stellar metallicity or the nebular oxygen abundance is high in comparison with other dwarf galaxies. The most massive object, Arp86a, has the expected metallicity and oxygen nebular abundance for a dwarf galaxy, while Arp72d has very large uncertainties in these values, so we cannot conclude one way or the other.

\subsection{Oxygen abundance gradients}

\begin{figure}
\centering
\includegraphics[width=0.9\linewidth]{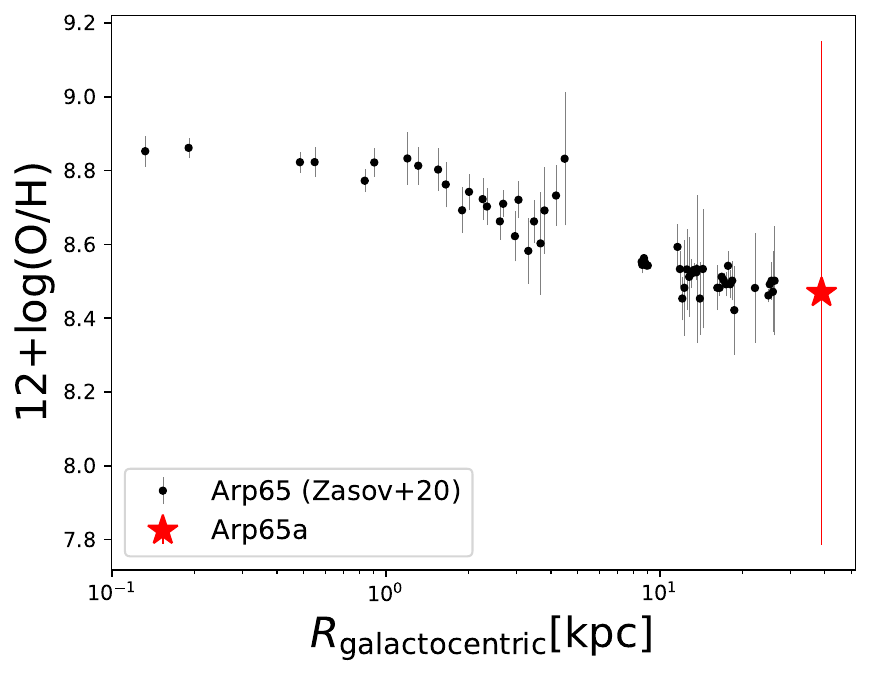} \\
\includegraphics[width=0.9\linewidth]{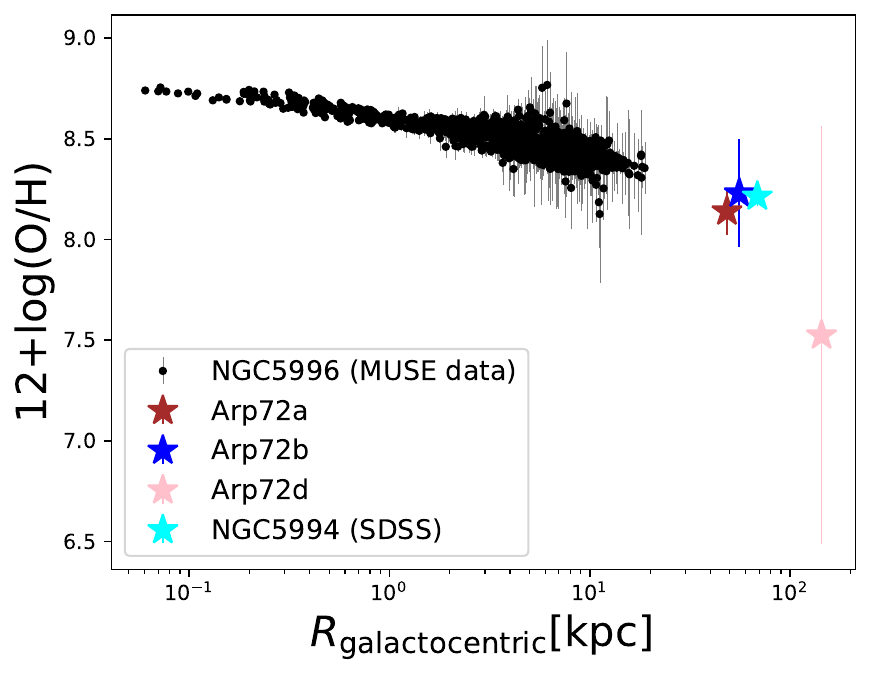} \\ 
\includegraphics[width=0.9\linewidth]{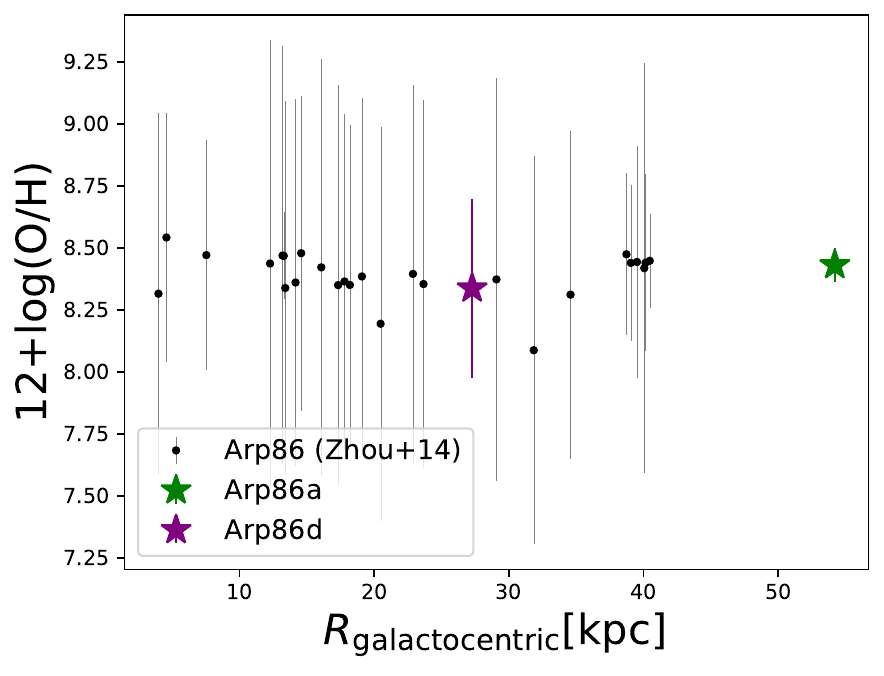} \\
\caption{\label{fig_ox_gradient} Oxygen abundance versus galactocentric radius for the six spectroscopically confirmed TDGs (colored star symbols). We show the nebular regions along the host galaxies as black dots and as a cyan star for NGC5994 (middle pannel).}
\end{figure}

 Another test of the TDG hypothesis is comparison with metallicities within the disks of the galaxies.
 We used the homogenized HyperLeda database\footnote{\url{http://leda.univ-lyon1.fr/}} to obtain the inclination of the main galaxies.
We took these inclinations
 into account when 
estimating the galactocentric radius for each spaxel in 
 the MUSE  data for Arp 72 and in each region reported in \cite{2014RAA....14.1393Z}  and \cite{2020MNRAS.498..101Z}. 
 For Arp 65, we assumed that NGC 90, the spiral to the west, is the parent galaxy.
The oxygen abundance gradients for each interacting system are plotted in Fig. \ref{fig_ox_gradient}, where the detached TDGs are shown as star symbols. All the TDGs exhibit an oxygen abundance consistent with the gradient of the main galaxy, that is, the measured abundance is approximately the expected one for the radii where the TDGs are located, except in the case of Arp86, where the oxygen abundances exhibit too much dispersion to visualize any gradient. However, the oxygen abundance values in Arp86 are within the observed range in the main galaxy and tails.  In the case of Arp72d, which is at a deprojected distance of 150 kpc, the oxygen abundance is lower than expected from the observed gradient in Arp72, although the uncertainty is large due to the faint detection of the [NII] emission line, so the oxygen abundance could be compatible with that of the disk.

\subsection{The effective radii of the tidal dwarf galaxies } \label{sec:sizes}
 
 \begin{figure}
\centering
\includegraphics[width=0.9\linewidth]{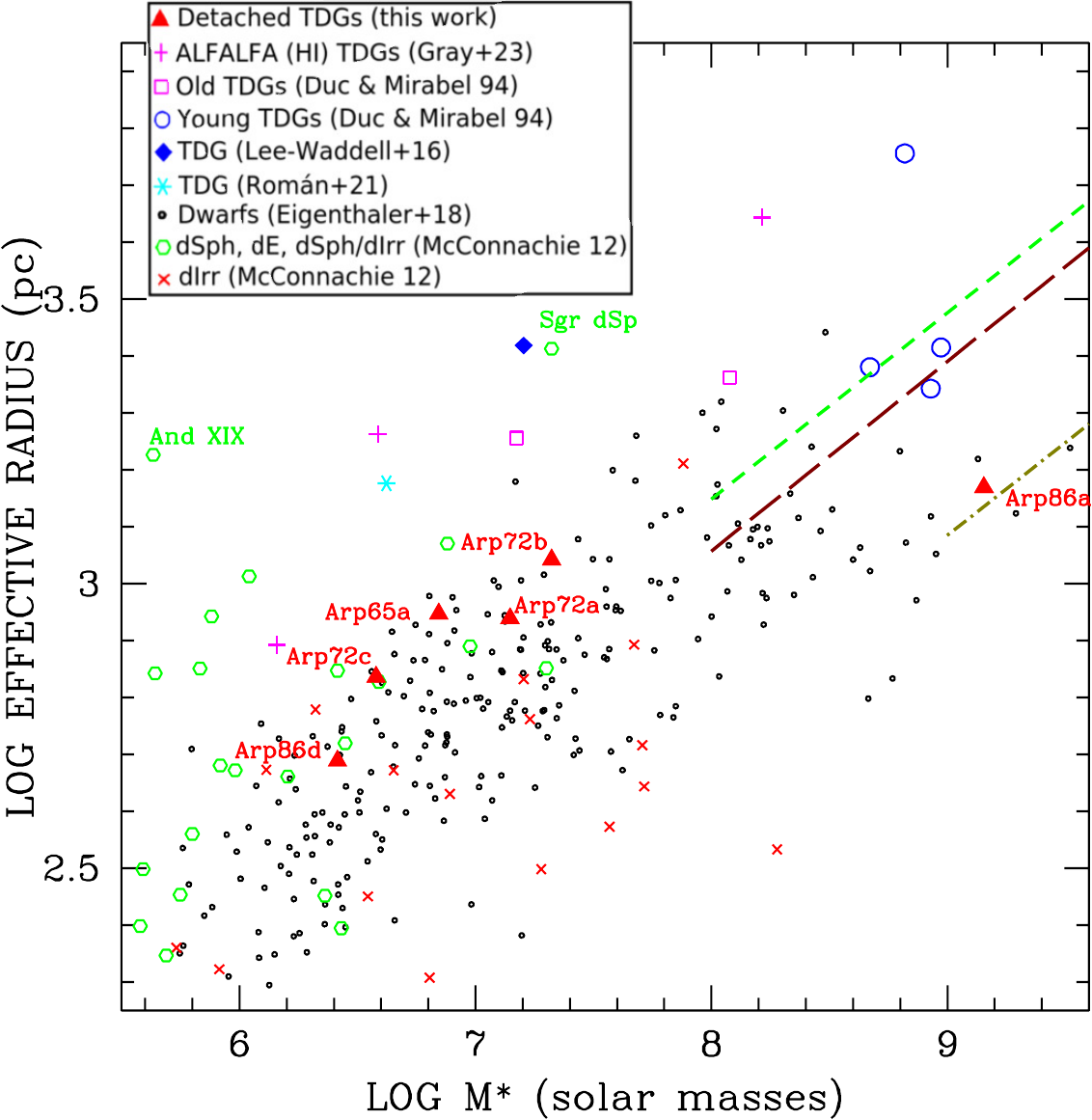} 
\caption{Effective radii versus stellar mass. 
Red filled triangles are our candidate TDGs. 
Magenta crosses are ALFALFA-selected candidate TDGs from \citet{gray2023}.
Magenta open squares are candidate old TDGs from \citet{duc1994};
Blue open circles are young TDGs in tidal features, as tabulated by \citet{duc1994}.
Blue filled diamond is the TDG AGC 208457, from \citet{lee2016}.
Cyan asterisk is the candidate TDG from \citet{2021A&A...649L..14R}.
Small black dots are dwarf galaxies in the Fornax cluster, from \citet{2018ApJ...855..142E}.
Small green open hexagons are Local Group  dSph, dE,  dSph/dIrr, from \citet{2012AJ....144....4M}. 
Red x marks are Local Group dwarf Irregular galaxies, from \citet{2012AJ....144....4M}.
The two
largest Local Group galaxies are labeled in green. Best effective radii-stellar mass fits from \citet{2016MNRAS.462.1470L} are shown as green dashed line (Sd-Irr galaxies), brown dashed line (Sab-Scd galaxies), and olive dot-dash line (S0-Sa galaxies).
\label{fig_effrad}}
\end{figure}
 
\cite{duc2014} suggest that
TDGs have larger effective radii relative to their stellar masses,
compared to other dwarf galaxies.
This means
that effective radii may be a powerful tool for identifying 
detached TDGs.  
The nearness of our sample galaxies and the availability of  DESI-LIS
images makes such an analysis possible. 
At the distance of our objects,
the median  DESI-LIS seeing of 1\farcs1 FWHM \citep{dey2019} corresponds to 0.3 kpc, and thus
regions that have diameters of 1 $-$ 2 kpc are well resolved.

 \begin{table}
\begin{center}
{\centering
\caption{\label{tab_radeff}  Effective radii in the r band of the detached TDG candidates. }
\begin{tabular}{cc}
Name & R$_{\rm{eff\thinspace r}}$\\
 & kpc   \\
\hline
Arp65a &  0.89  \\
Arp72a & 0.87  \\
Arp72b & 1.10  \\
Arp72c & 0.69  \\
Arp72d & 1.42  \\
Arp86a & 1.48  \\
Arp86d & 0.49  \\
\end{tabular}
}
\end{center}
 
\end{table}  
 
We used the  DESI-LIS r images to determine the effective radii (half-light radii) of the candidate TDGs.
These radii are provided in Table \ref{tab_radeff}.   In Figure \ref{fig_effrad}, we plot these effective radii (red filled triangles)
against the stellar masses, and compare with effective radii of other candidate TDGs from earlier studies
\citep{duc2014,lee2016,2021A&A...649L..14R,gray2023}. 
In Figure \ref{fig_effrad}, we also plot Local Group dwarf Irregular galaxies  (dIrr, green open circles) and  dwarf spheroidal  (dSph) or 
dSph/dIrr galaxies from \cite{2012AJ....144....4M} and the \cite{2018ApJ...855..142E} Fornax dwarf galaxies. 
 We also show the best effective radii-stellar mass fits from \citet{2016MNRAS.462.1470L} as green dashed line (Sd-Irr galaxies), brown dashed line (Sab-Scd galaxies), and olive dot-dash line (S0-Sa galaxies).

Figure \ref{fig_effrad} shows that Arp 86a is relatively compact for 
its stellar mass respect to other TDGs, and is similar in size with S0-Sa galaxies from \citet{2016MNRAS.462.1470L} (olive line), consistent with our conclusion based on metallicity that it is
probably a preexisting dwarf galaxy rather than a TDG. 
In contrast, 
our other candidate TDGs are larger than most Local Group dIrr galaxies with the same 
stellar mass, are more similar in size to Local Group  dSph galaxies with similar masses, and lie above the extrapolation of the Sd-Irr relation from \citet{2016MNRAS.462.1470L} to low masses (green line).
For a given stellar mass, the TDGs in our sample lie at the upper edge of the size 
range of Fornax dwarf galaxies.   The five confirmed TDGs can be split into two groups of stellar mass: one that is less massive within the $\rm{log}(M_{*})=[6.2,6.6]$ range (Arp65a, Arp72c, and Arp86d) and another that is more massive within the $\rm{log}(M_{*})=[7.1,7.2]$ range (Arp72a and Arp72b). 
In the less-massive range, the average effective radius of dwarf galaxies from \citet{2018ApJ...855..142E} is $370\pm 50 \rm{pc}$, while the effective radius of the less-massive detached TDGs from this work are larger (see Table \ref{tab_radeff}).  
In the more-massive range, the average effective radius of dwarf galaxies from \citet{2018ApJ...855..142E} is $680\pm 110 \rm{pc}$, while the sizes of the more-massive detached TDGs  from this work are also larger (see Table \ref{tab_radeff}).

The confirmed TDGs in this work, with stellar masses $M_{*}<1.7\times10^7\rm{M_{\odot}}$, are less massive
than most previously known TDGs; only
three of the twelve previously identified TDGs in
Figure \ref{fig_effrad}  have masses $M_{*}<10^7\rm{M_{\odot}}$.
Our confirmed TDGs have effective radii $R_{\rm{eff}}<1.10$kpc, so they are smaller than the previous identified TDGs, except one detected from the  Arecibo Legacy Fast ALFA (ALFALFA) survey.

\section{Discussion}

 The spectral resolution of the observations, $4.4\rm{\AA}$, corresponds to a velocity resolution of about 200 km~s$^{-1}$ at H$\alpha$.
This is too large to allow for reliable measurement of the 
rotation curves of these TDGs; thus, we could not directly measure the dynamical masses.
That means we cannot test directly for the presence (or absence) of Dark Matter in the TDGs.
However, we can address this issue indirectly, via the timescale of star formation triggering.
According to simulations by \cite{lelli2015}, 
dynamical equilibrium is achieved 
in gas-dominated TDGs within 200 Myrs.
This means that 
all six of our detached TDGs should be either in or very close to dynamical equilibrium, 
assuming that the TDGs formed simultaneously with the star formation burst.
If  the TDGs are in dynamical equilibrium, the dynamical mass should be equal to the total mass (baryonic plus Dark Matter).

The survival timescale of TDGs is crucial for understanding their implications in galaxy evolution and interpreting observations in   
 terms of dark galaxies 
\citep{2021A&A...649L..14R}. Our burst ages indicate that  TDGs can survive for at least 900 Myr, as observed in the case of Arp72b. 
 We note that this represents a lower 
limit  to the possible ages of TDGs since our observations are biased toward emission line-emitting objects
 because we used the H$\alpha$ line for redshift confirmation.
It is  possible that TDGs can survive even longer than what we have observed in this sample.

 \citet{2019A&A...626A..47H} reported that gas-rich TDGs from the Illustris-1 simulation do not have Dark Matter, are gravitationally bound, and are in virial equilibrium. Our TDGs, selected because they show star formation and nebular emission lines, are gas-rich. Consequently, they should be compared with the gas-rich TDGs from  \citet[][sample B]{2019A&A...626A..47H}. Their  stellar
 masses and ages are close to those in Sample B, which have median values of $3 \times 10^6~\rm{M_{\odot}}$ and $1$Gyr.  In the Illustris-1 simulation, a "stellar mass particle" has a mass of 1.26 $\times 10^6~\rm{M_{\odot}}$, and thus such low-mass TDGs are not well resolved in the simulation. 
 \citet{2018MNRAS.474..580P} reported gas-rich TDG candidates and Dark Matter-free objects detected in the Evolution and Assembly of GaLaxies and their Environments (EAGLE) cosmological simulation, having a mean stellar mass of $2 \times 10^6~\rm{M_{\odot}}$, also in agreement with the values we find in this study.  A "stellar mass particle" in the EAGLE simulation is 
 2.26 $\times$ 10$^5~\rm{M_{\odot}}$. Although TDGs do not have Dark Matter, the use of gas dynamics to test different cosmological scenarios can be problematic if the gas is not in virial equilibrium \citep{2016MNRAS.457L..14F}. The best TDGs to examine for the presence and effects of Dark Matter are those from Arp72, as they are old enough to be in equilibrium.

\subsection{Comparison to other studies} \label{sec:other}


To investigate the longevity of TDGs, \cite{bournaud2006} 
ran 96 N-body simulations of galaxy interactions with a range of 
mass ratios and orbital parameters. About 75\% of the TDGs 
formed in these simulations either 
fell back onto their parent galaxies or became disrupted within
300 Myrs. According to their models,
long-lasting TDGs are more likely to form in prograde
encounters between approximately equal mass galaxies.
\cite{bournaud2006} concluded that pairs with mass ratios 
between 1/4 and 8 are the most favorable for the production of
long-lived TDGs.

Two out of the three systems in which we find detached TDGs,
Arp 72 and Arp 86, have companions with lower masses than expected
based on these models. If 3.6 $\mu$m flux ratio is that of the stellar massz, 
the mass ratio of Arp72 (1/12) is rather different from what is reported by 
\cite{bournaud2006}, which suggests that the detached TDGs from Arp 72 are
probably different from those studied in the simulations of  \citet{bournaud2006}.

Previous studies have tried to discover detached TDGs. 
Using optical and HI imaging, \cite{duc2014}
obtained
optical spectra of seven
dwarf galaxy-like objects close to nearby early-type galaxies.
If these early-type galaxies are the relics of galaxy mergers,
some of these dwarfs may have been formed tidally.
One of their candidate TDGs
appears to be connected to a nearby
elliptical galaxy by a faint tidal tail, and it has a 
higher metallicity than
expected given its stellar mass, suggesting a tidal origin.

\cite{delgado2003}
obtained deep optical
images of six merger remnants, searching for possible detached
TDGs in their vicinities.  Their statistics were consistent
with expected number counts of background
galaxies, and they concluded
that there is no evidence for
a large population of locally formed dwarf galaxies
near those galaxies.
More recently, \citet{gray2023} confirmed two detached TDGs based on HI observations from 
the ALFALFA survey and deep optical imaging. 

Detached TDGs do not seem to be very common, although they could be more frequent but difficult to detect due to their brightness and the limited spatial field of view of spectroscopic and optical observations. Wide surveys will help improve the spatial coverage of nearby galaxy mergers, where detached TDG candidates can be detected. However, spectroscopic observations would still be needed to confirm the membership of the TDG candidates to galaxy mergers. Wide surveys such as Javalambre Physics of the Accelerating Universe \citep[J-PAS,][]{2014arXiv1403.5237B,2021A&A...653A..31B} will cover  8500$\deg^2$ in 54 narrowband filters with a pseudo spectral resolution of $R\sim40$, where redshift confirmation will be possible for detached TDG characterization studies thereafter. Therefore, we will be able to improve the detections of these detached TDGs in order to have a complete census from which to estimate formation and survival rates.  
In turn, this will allow us to understand the impact of TDGs on galaxy evolution and dwarf galaxy formation.  It will also allow us to identify the best candidates for detailed follow-up studies. The Baryonic Tully-Fisher Relation (BTFR), which relates the baryonic mass with the rotation velocity of galaxies, has been postulated as a fundamental relation \citep{2000ApJ...533L..99M,2005ApJ...632..859M,2019MNRAS.484.3267L}. However, due to the complexity of dynamical modeling of faint dwarf galaxies with turbulent gas, the origin of a possible deviation from the BTFR relation of TDGs (or other dwarf galaxies) 
is still an open question \citep{lelli2015,2020NatAs...4..246G,2024ApJ...964...85D}.

Detailed dynamical analysis of detached TDGs could be
used to test the validity of the BTFR for TDGs. Since they are already detached from the main galaxies, they should be simpler to analyze from a dynamical perspective and thus can be used to test the different predictions for the BTFR from various theories \citep{2019MNRAS.484.3267L}. The presence of any dynamically measured amount of nonvisible matter in TDGs has implications for the distribution of dark matter or the existence of dark gas \citep{lelli2015}.

Stellar feedback is commonly invoked as one of the drivers of stellar mass-growth regulation in the $\Lambda$ Cold Dark Matter ($\Lambda$-CDM) standard cosmological model \citep{2012RAA....12..917S}. Dark matter deficient galaxies such as TDGs should experience stronger stellar feedback due to the lack of mass compared to dark matter dominated dwarf galaxies. Detailed analyses of detached TDGs, such as those using chemical enrichment histories \citep{2013ApJ...772..119L,2018ApJ...852...74B} or star formation histories \citep{2019MNRAS.487L..61Z,2020MNRAS.499.1172Z}, can impose constraints on the stellar feedback needed to match the predicted relations from $\Lambda$-CDM galaxy simulations.

\section{Conclusions}

In our search for detached TDGs associated with a sample of 39 interacting galaxies, we spectroscopically confirmed the redshifts of seven detached TDG candidates. For six
of these objects, we utilized the Boller and Chivens long-slit spectrograph at the 2.1m telescope at the San Pedro Mártir Observatory. 
We used an SDSS spectrum for the seventh source. 
Nebular oxygen abundances were derived using the S-calibrator from \citet{Pilyugin16}, and stellar metallicities, masses, and recent burst ages were determined through SED fitting using CIGALE  \citep{2019A&A...622A.103B}. Based on the fundamental stellar and gas metallicity-stellar mass relations, we confirmed the tidal nature of five out of the seven detached TDG candidates. One candidate, Arp72d, remains uncertain due to high uncertainties in nebular abundance and stellar metallicity, while the most massive one, Arp86a, is compatible with being a preexisting dwarf galaxy.

 We compared the nebular oxygen abundances of the seven TDG
candidates with the 
(O/H)-to-galactocentric radius relation of the host galaxies.
The TDG candidates exhibit oxygen abundances consistent with the observed abundance gradients in the host galaxies. The effective radii-stellar mass relation of these detached TDGs indicates that they tend to be larger than typical dwarf galaxies on average, while they are smaller than other TDGs for a given mass. The most massive TDG candidate, Arp86a,
is more compact relative to its mass and is consistent in size with other dwarf
galaxies of its mass, supporting the idea that it is a 
preexisting dwarf galaxy rather than a TDG.

Two of the interacting galaxy systems, Arp72 and Arp86, have merger ratios of approximately 1/10 and 1/5, respectively. This contrasts with N-body simulations \citep{bournaud2006}, which suggest that detached TDGs have a higher survival probability for merger ratios between 1/4 and 8/1, indicating potential differences in formation and survival mechanisms.

The detection and study of detached TDGs has significant implications for dwarf galaxy formation and the discovery of so-called dark galaxies,  that is, dark matter halos that were unable to transform their gas into stars. The optical brightnesses of TDGs decreases with time, so if detached TDGs are frequent, have large survival times, and become gravitationally unbound from the main galaxies, it is probable that many identified as dark galaxies are indeed detached TDGs. However, limitations in depth and field of view of observations raise questions about the low number of detections, whether due to actual scarcity or observational factors. The confirmed detached TDGs presented in this work offer valuable opportunities for studying TDG evolution, star formation in extreme environments, dwarf galaxy formation, and the testing of $\Lambda$-CDM predictions regarding the BTFR, stellar feedback, and the distribution of dark matter.

\begin{acknowledgements} 
 
Based upon observations carried out at the Observatorio Astronómico Nacional on the Sierra San Pedro Mártir (OAN-SPM), Baja California, México. We thank the daytime and night support staff at the OAN-SPM for facilitating and helping obtain our observations. JZC acknowledges the financial support provided by the Instituto Nacional de Astrof\'isica, \'Optica y Electr\'onica, and the Governments of Spain and Arag\'on through their general budgets and the Fondo de Inversiones de Teruel. BJS and MLG were supported by National Science Foundation Extragalactic Astronomy grant ASTR-1714491 and from the NASA Tennessee Space Grant. We would like to express our gratitude to the anonymous referee for their valuable comments and suggestions, which have significantly improved the quality of this paper. Their thorough review and constructive feedback were greatly appreciated.

This research has made use of the NASA/IPAC Extragalatic Database (NED), which is operated by the Jet Propulsion Laboratory, California Institute of Technology, under contract with NASA. This work is based in part on observations made with the \textit{Spitzer} Space Telescope, which is operated by the Jet Propulsion Laboratory (JPL), California Institute of Technology under a contract with NASA. This study also uses data from the NASA Galaxy Evolution Explorer (GALEX), which was operated for NASA by the California Institute of Technology under NASA contract NAS5-98034. We acknowledge the usage of the HyperLeda database (\url{http://leda.univ-lyon1.fr})

Based on observations collected at the European Southern Observatory under ESO programme 0101.D-0748(B) and data obtained from the ESO Science Archive Facility with DOI under https://doi.org/10.18727/archive/41.

The Legacy Surveys consist of three individual and complementary projects: the Dark Energy Camera Legacy Survey (DECaLS; Proposal ID \#2014B-0404; PIs: David Schlegel and Arjun Dey), the Beijing-Arizona Sky Survey (BASS; NOAO Prop. ID \#2015A-0801; PIs: Zhou Xu and Xiaohui Fan), and the Mayall z-band Legacy Survey (MzLS; Prop. ID \#2016A-0453; PI: Arjun Dey). DECaLS, BASS and MzLS together include data obtained, respectively, at the Blanco telescope, Cerro Tololo Inter-American Observatory, NSF's NOIRLab; the Bok telescope, Steward Observatory, University of Arizona; and the Mayall telescope, Kitt Peak National Observatory, NOIRLab. Pipeline processing and analyses of the data were supported by NOIRLab and the Lawrence Berkeley National Laboratory (LBNL). The Legacy Surveys project is honored to be permitted to conduct astronomical research on Iolkam Du'ag (Kitt Peak), a mountain with particular significance to the Tohono O'odham Nation.

NOIRLab is operated by the Association of Universities for Research in Astronomy (AURA) under a cooperative agreement with the National Science Foundation. LBNL is managed by the Regents of the University of California under contract to the U.S. Department of Energy.

This project used data obtained with the Dark Energy Camera (DECam), which was constructed by the Dark Energy Survey (DES) collaboration. Funding for the DES Projects has been provided by the U.S. Department of Energy, the U.S. National Science Foundation, the Ministry of Science and Education of Spain, the Science and Technology Facilities Council of the United Kingdom, the Higher Education Funding Council for England, the National Center for Supercomputing Applications at the University of Illinois at Urbana-Champaign, the Kavli Institute of Cosmological Physics at the University of Chicago, Center for Cosmology and Astro-Particle Physics at the Ohio State University, the Mitchell Institute for Fundamental Physics and Astronomy at Texas A\&M University, Financiadora de Estudos e Projetos, Fundacao Carlos Chagas Filho de Amparo, Financiadora de Estudos e Projetos, Fundacao Carlos Chagas Filho de Amparo a Pesquisa do Estado do Rio de Janeiro, Conselho Nacional de Desenvolvimento Cientifico e Tecnologico and the Ministerio da Ciencia, Tecnologia e Inovacao, the Deutsche Forschungsgemeinschaft and the Collaborating Institutions in the Dark Energy Survey. The Collaborating Institutions are Argonne National Laboratory, the University of California at Santa Cruz, the University of Cambridge, Centro de Investigaciones Energeticas, Medioambientales y Tecnologicas-Madrid, the University of Chicago, University College London, the DES-Brazil Consortium, the University of Edinburgh, the Eidgenossische Technische Hochschule (ETH) Zurich, Fermi National Accelerator Laboratory, the University of Illinois at Urbana-Champaign, the Institut de Ciencies de l'Espai (IEEC/CSIC), the Institut de Fisica d'Altes Energies, Lawrence Berkeley National Laboratory, the Ludwig Maximilians Universitat Munchen and the associated Excellence Cluster Universe, the University of Michigan, NSF's NOIRLab, the University of Nottingham, the Ohio State University, the University of Pennsylvania, the University of Portsmouth, SLAC National Accelerator Laboratory, Stanford University, the University of Sussex, and Texas A\&M University.

BASS is a key project of the Telescope Access Program (TAP), which has been funded by the National Astronomical Observatories of China, the Chinese Academy of Sciences (the Strategic Priority Research Program ``The Emergence of Cosmological Structures'' Grant \# XDB09000000), and the Special Fund for Astronomy from the Ministry of Finance. The BASS is also supported by the External Cooperation Program of Chinese Academy of Sciences (Grant \# 114A11KYSB20160057), and Chinese National Natural Science Foundation (Grant \# 12120101003, \# 11433005).

The Legacy Survey team makes use of data products from the Near-Earth Object Wide-field Infrared Survey Explorer (NEOWISE), which is a project of the Jet Propulsion Laboratory/California Institute of Technology. NEOWISE is funded by the National Aeronautics and Space Administration.

The Legacy Surveys imaging of the DESI footprint is supported by the Director, Office of Science, Office of High Energy Physics of the U.S. Department of Energy under Contract No. DE-AC02-05CH1123, by the National Energy Research Scientific Computing Center, a DOE Office of Science User Facility under the same contract; and by the U.S. National Science Foundation, Division of Astronomical Sciences under Contract No. AST-0950945 to NOAO.

Funding for the Sloan Digital Sky Survey IV has been provided by the Alfred P. Sloan Foundation, the U.S. Department of Energy Office of Science, and the Participating Institutions. SDSS acknowledges support and resources from the Center for High-Performance Computing at the University of Utah. The SDSS web site is www.sdss4.org.

SDSS is managed by the Astrophysical Research Consortium for the Participating Institutions of the SDSS Collaboration including the Brazilian Participation Group, the Carnegie Institution for Science, Carnegie Mellon University, Center for Astrophysics - Harvard \& Smithsonian (CfA), the Chilean Participation Group, the French Participation Group, Instituto de Astrofísica de Canarias, The Johns Hopkins University, Kavli Institute for the Physics and Mathematics of the Universe (IPMU) / University of Tokyo, the Korean Participation Group, Lawrence Berkeley National Laboratory, Leibniz Institut für Astrophysik Potsdam (AIP), Max-Planck-Institut für Astronomie (MPIA Heidelberg), Max-Planck-Institut für Astrophysik (MPA Garching), Max-Planck-Institut für Extraterrestrische Physik (MPE), National Astronomical Observatories of China, New Mexico State University, New York University, University of Notre Dame, Observatório Nacional / MCTI, The Ohio State University, Pennsylvania State University, Shanghai Astronomical Observatory, United Kingdom Participation Group, Universidad Nacional Autónoma de México, University of Arizona, University of Colorado Boulder, University of Oxford, University of Portsmouth, University of Utah, University of Virginia, University of Washington, University of Wisconsin, Vanderbilt University, and Yale University.

\end{acknowledgements}

\bibliographystyle{aa}

\end{document}